\documentclass[twocolumn]{aastex631}

\usepackage{amsmath}
\usepackage{mathtools}
\usepackage{multirow}
\usepackage{tabularx}
\usepackage{natbib}

\begin{document}

\title{Lomb-Scargle periodograms struggle with non-sinusoidal supermassive BH binary signatures in quasar lightcurves}

\author{Allison Lin}
\affiliation{Department of Astronomy, Columbia University, New York, NY 10027, USA}

\author[0000-0003-3579-2522]{Maria Charisi} 
\affiliation{Department of Physics and Astronomy, Washington State University, Pullman, WA 99163, USA}
\affiliation{Institute of Astrophysics, FORTH, GR-71110, Heraklion, Greece}

\author[0000-0003-3633-5403]{Zolt\'an Haiman}
\affiliation{Department of Astronomy, Columbia University, New York, NY 10027, USA}
\affiliation{Department of Physics, Columbia University, New York, NY 10027, USA}
\affiliation{Institute of Science and Technology Austria (ISTA), Am Campus 1, Klosterneuburg 3400, Austria}

\begin{abstract}

Supermassive black hole binary (SMBHB) systems are expected to form as a consequence of galaxy mergers. At sub-parsec separations, SMBHBs can be identified as quasars with periodic variability with previous periodicity searches uncovering significant candidates. However, these searches focused primarily on sinusoidal signals, while theoretical models and hydrodynamical simulations predict that binaries produce more complex non-sinusoidal pulse shapes. Here we examine the efficacy of the Lomb-Scargle periodogram (LSP; one of the most popular tools for periodicity searches in unevenly sampled lightcurves) to detect periodicities with a sawtooth shape mimicking results of hydrodynamical simulations. We simulate idealised well-sampled lightcurves, lightcurves that mimic the data in the Palomar Transient Factory (PTF) analyzed in \cite{Charisi2016}, and lightcurves that resemble our expectations for single-band data in the upcoming Legacy Survey of Space and Time (LSST) of the Rubin Observatory. We approximate quasar variability with a damped random walk (DRW) model, inject sinusoidal and sawtooth pulse shapes and assess their statistical significance.
We find that in the presence of red noise the LSP detects a relatively low fraction of the sinusoidal signals ($\sim$45\%, $\sim$24\% and $\sim$23\%, in the PTF-like, idealised, and LSST-like lightcurves, respectively). The fraction is significantly reduced for sawtooth periodicity (with only $\sim$9\% in PTF-like and 
$\sim$1\% in idealised and LSST-like lightcurves). These low recovery rates imply that previous searches have missed the large majority of binaries. They also have significant implications for the detection of SMBHBs in upcoming LSST necessating the developement of advanced tools that go beyond the simple LSP.

\end{abstract}


\section{Introduction}
\label{sec:intro}

Observational evidence suggests that supermassive black holes (SMBHs) exist at the center of every massive galaxy \citep{Richstone1998,KormendyHo2013}. The SMBHs play an important role in shaping their host galaxies, since the mass of the SMBHs is correlated to several properties of the host. 
In addition, frequent mergers of galaxies naturally deliver two SMBHs to the nucleus of the post-merger galaxies, which can eventually form bound SMBH binaries \citep{Begelman1980}. After the galaxies merge, the SMBHs sink towards the center of the post-merger galaxy through dynamical friction. As the SMBHs spiral closer toward each other, they form a near-Keplerian system. When they reach separations of a few parsecs, gravitational interactions with nearby stars and the gaseous circumbinary disk can continue to shrink the binary orbits.
If these mechanisms are sufficiently effective to overcome the 'final-parsec problem,' gravitational wave (GW) emission takes over, accelerating the inspiral toward the merger. 

The most massive systems can be detected at nanohertz GW frequencies with pulsar timing arrays (PTAs)
\citep{Burke-Spolaor2019,2021arXiv210513270T}. In fact, all major PTA collaborations,
like the North American Nanohertz Observatory for Gravitational Waves (NANOGrav), the European Pulsar Timing Array (EPTA) along with the Indian Pulsar Timing Array (InPTA), the Parkes Pulsar Timing Array (PPTA), the Chinese Pulsar Timing Array (CPTA) and the MeerKAT Pulsar Timing Array (MPTA) have recently detected evidence for a GW background \citep{2023ApJ...951L...8A, 2023A&A...678A..50E, 2023ApJ...951L...6R, 2023RAA....23g5024X,MPTA}, which is likely produced by a population of SMBH binaries \citep{2023ApJ...952L..37A, 2024A&A...685A..94E}. Individual binaries are expected to be resolved on top of this background in a few years \citep{Kelley2018, 2020PhRvD.102h4039T, 2022arXiv220701607B}. 

In addition to strong GWs, binaries are also expected to produce bright electromagnetic (EM) emission, since
SMBH binaries reach small separations surrounded by large amounts of gas, which accretes to the SMBHs and produces bright EM signatures. Therefore,  binaries are promising sources for multi-messenger observations combining EM+GW data \citep{Kocsis+2006,2009ApJ...700.1952H,2019BAAS...51c.490K, 2022MNRAS.510.5929C}.

Given that binary systems are expected to form in gas-rich environments, several hydrodynamic simulations have explored such systems embedded in gaseous disks. These studies suggest that binaries may manifest as quasars with periodic variability.
More specifically, simulations demonstrate that torques exerted by the binary expel the gas from the central-most region and create a low-density cavity. As the orbital motion of the binary perturbs the cavity edge, the SMBHs pull streams of gas inwards, leading to periodic mass transfer from the disk onto the binary components, which translates into periodic brightness fluctuations~\citep[e.g.][]{westernacher-schneider_multiband_2022}. Periodic variability can also arise due to Doppler boost from gas bound to the individual SMBHs, and from self-lensing effects (see \citealt{2021arXiv210903262B} and \citealt{Dan&Charisi2023} for recent reviews). 

Many observational studies have searched for this EM signature in time-domain surveys and identified several candidate SMBHBs as quasars with periodically varying optical emission. 
Note that this approach associates the bright quasar phases with SMBHBs, which is justified as quasars are thought often to be triggered by mergers \citep{KauffmanHaehnelt2000}. However, the search requires large surveys, because 
the binary lifetime due to gravitational-wave inspiral would be of order $10^5$ years, which is $\sim10^{-3}$ of the typical expected bright quasar phase~\citep{2009ApJ...700.1952H}. This implies that even if most quasars are binaries, we would expect to find them only in a small fraction $\lesssim 10^{-3}$ of quasars.  Suitably large
systematic searches in the Catalina Real-Time Transient Survey (CRTS; \citealt{Graham2015}), the Palomar Transient Factory (PTF; \citealt{Charisi2016}), 
the Panoramic Survey Telescope and Rapid Response System (PanSTARRS; \citealt{2019ApJ...884...36L}), the Dark Energy Survey (DES; \citealt{Chen2020}), the Zwicky Transient Facility (ZTF; \citealt{Chen_2024_ztf}), the Wide-field Infrared Survey Explorer (WISE; \citealt{2025ApJ...978...86L}) and Gaia \citep{2025arXiv250516884H} have revealed over 200 candidates to date. Additional candidates were also identified individually in lightcurves from long-term observing campaigns \citep{2006IJMPD..15..261D, 2016ApJS..225...29B, 2019ApJS..241...33L}.

However, quasars exhibit red noise variability, which can mimic periodic signals introducing false positives, while at the same time hindering the identification of genuine periodicity \citep{Vaughan2016,Witt2022,2024ApJ...965...34D,Robnik2024}.
This is exacerbated by limitations in our understanding of the stochastic variability of quasars, which in previous periodicity searches was modeled with a DRW model \citep{Kelly2009,MacLeod+2010, Koz2010, 2021ApJ...907...96S, Burke_2021} and the fact that binaries typically have relatively long periods allowing only 2-3 cycles to be observed \citep{Vaughan2016}. Therefore, additional signatures, such as spectral line shifts indicative of radial motion \citep{eracleous12}, multi-wavelength Doppler boost \citep{Dorazio2015Nature,2018MNRAS.476.4617C,2020MNRAS.496.1683X}, distorted radio jets \citep{1993ApJ...409..130R,2013MNRAS.428..280C} or atypical X-ray spectra \citep{2020ApJ...900..148S}, are required to boost the significance of each candidate.
We refer the reader to \citet{DeRosa2019} and \citet{Dan&Charisi2023} for reviews on binary signatures and observational searches.

To date, most searches for periodic variability have focused on sinusoidal periodicity, which is the simplest periodic signal.  A widely used method for detecting periodic signals in unevenly sampled data is the Lomb-Scargle periodogram (LSP; \citealt{lomb_least-squares_1976,scargle_studies_1982} or \citealt{vanderplas_understanding_2018} for a recent review). The LSP (along with other Fourier-based techniques) is particularly effective in identifying sinusoidal variations, but may be less sensitive to signals that deviate from this shape. Such signals can be common in binaries, e.g., when the periodicity arises from periodic accretion or the binary is Doppler-modulated along an eccentric orbit \citep{2020MNRAS.495.4061H, duffell_circumbinary_2020, 2021ApJ...909L..13Z,westernacher-schneider_multiband_2022}. In addition, the LSP may be a sub-optimal statistic for periodic signals in the presence of non-white stochastic noise, like in quasars \citep{Robnik2024}.

In this paper, we investigate the ability of the LSP to detect non-sinusoidal periodicity in quasar lightcurves under different survey conditions, focusing specifically on sawtooth periodic patterns, which approximate the pulse shapes predicted by hydrodynamic simulations~\citep{duffell_circumbinary_2020,westernacher-schneider_multiband_2022}. To explore this effect, we generate idealised and realistic PTF-like mock lightcurves, exhibiting sawtooth periodic signals, with realistic quasar noise modeled as a DRW, as in previous quasar periodicity searches. Since the upcoming LSST of the Rubin Observatory is expected to play a significant role in quasar periodicity searches \citep{2021MNRAS.506.2408X,2021MNRAS.508.2524K,2024A&A...691A.250C}, we also examine single-band LSST-like lightcurves. We also emphasize that the underlying variability of quasars is modeled with a DRW to reflect the common practice in previous periodicity searches, but this model does not fully capture quasar variability and should be viewed as an approximation of quasar noise.
For comparison, we repeat the analysis for sinusoidal signals with similar properties.

This paper is outlined as follows: In Section~\ref{sec:methods}, we describe the mock quasar lightcurves and the statistical methods used to determine the significance of the periodicity. In Section~\ref{sec:results}, we report the fraction of detected periodic (sinusoidal and sawtooth) signals in both idealised, PTF-like and LSST-like lightcurves. In Section~\ref{sec:discuss}, we discuss our findings, and in Section~\ref{sec:conclusion}, we summarise our results and offer our conclusions.

\section{Methods}
\label{sec:methods}

Our objective is to study the recovery of sawtooth periodicity and compare it with the recovery of sinusoidal signals with similar periods and amplitudes under different survey conditions. We explore the detection of such signals in three cases: highly idealised lightcurves with continuous regular sampling, lightcurves that resemble the ones observed in PTF, and lightcurves that represent our expectations for single-band well-sampled LSST lightcurves, which is expected to play a crucial role in quasar periodicity searches in the upcoming years. For the detection, we follow the statistical analysis in \cite{Charisi2016}, which, like nearly all similar periodicity searches, is based on the LSP. 
Below we describe in detail our quasar lightcurve simulations and our methodology for detecting periodic signals and assessing their statistical significance. 

\begin{figure}
    \centering
    \includegraphics[width=0.45\textwidth]{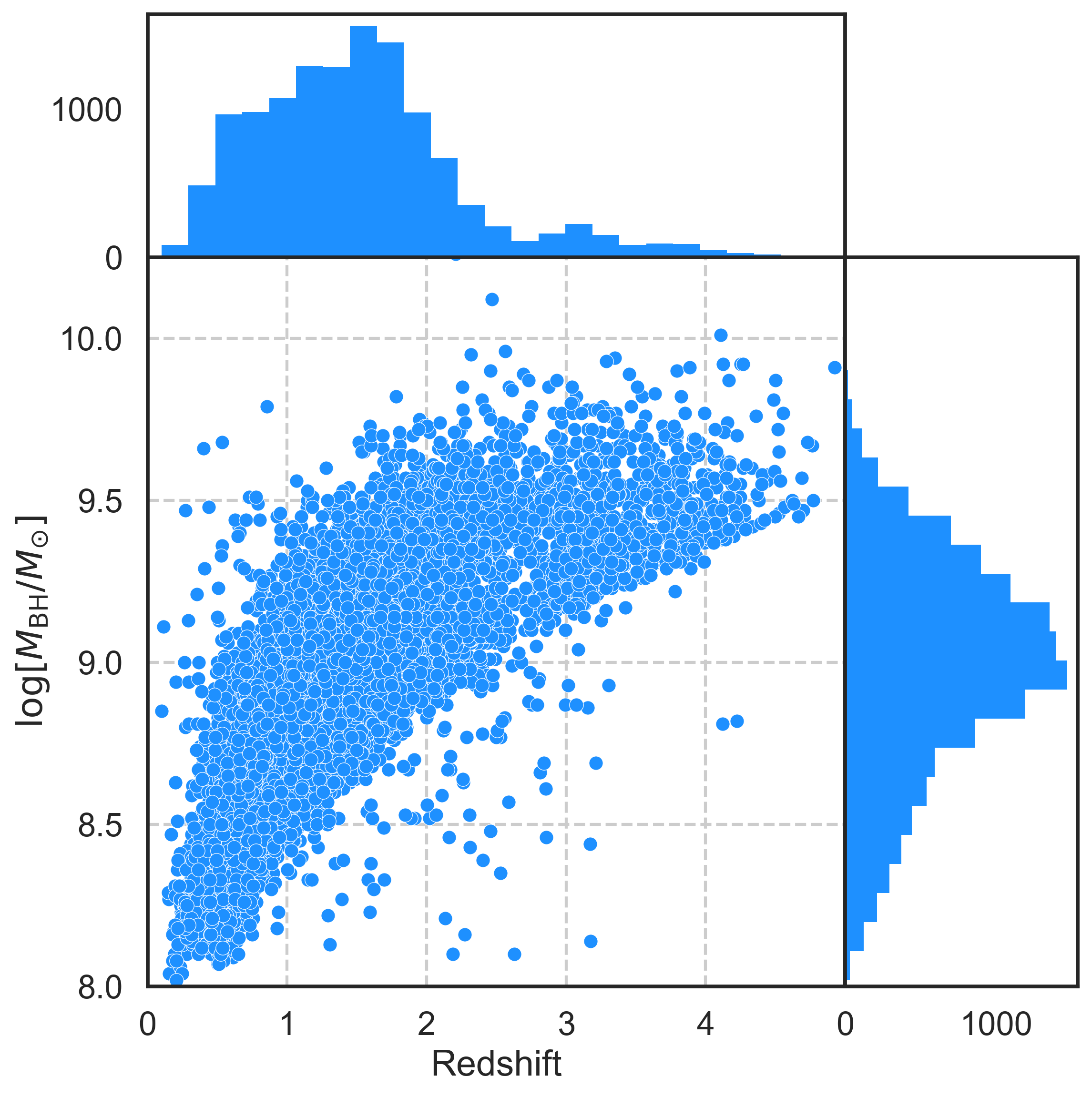}
    \caption{SMBH mass versus redshift for the 12,400 quasars in our sample. The marginalised distributions of the two quantities are also shown. The quasars in our sample span a wide range in SMBH mass and redshift, representative of the entire quasar population.}
    \label{Fig:Mass_Redshift}
\end{figure}

\subsection{Quasar sample}
\label{sec:sample}
In order to evaluate the ability of the LSP to detect non-sinusoidal signals in a realistic setting, we examine the sample of quasars detected by PTF and analysed in \cite{Charisi2016}. By closely following their statistical analysis for periodicity detection, this study provides one additional benefit; it allows us to assess whether non-sinusoidal periodicity would have been detected by the pipeline if present in the dataset, or such signals would have been missed and to what extent.

\begin{figure*}
    \centering
    \includegraphics[width=0.45\textwidth]{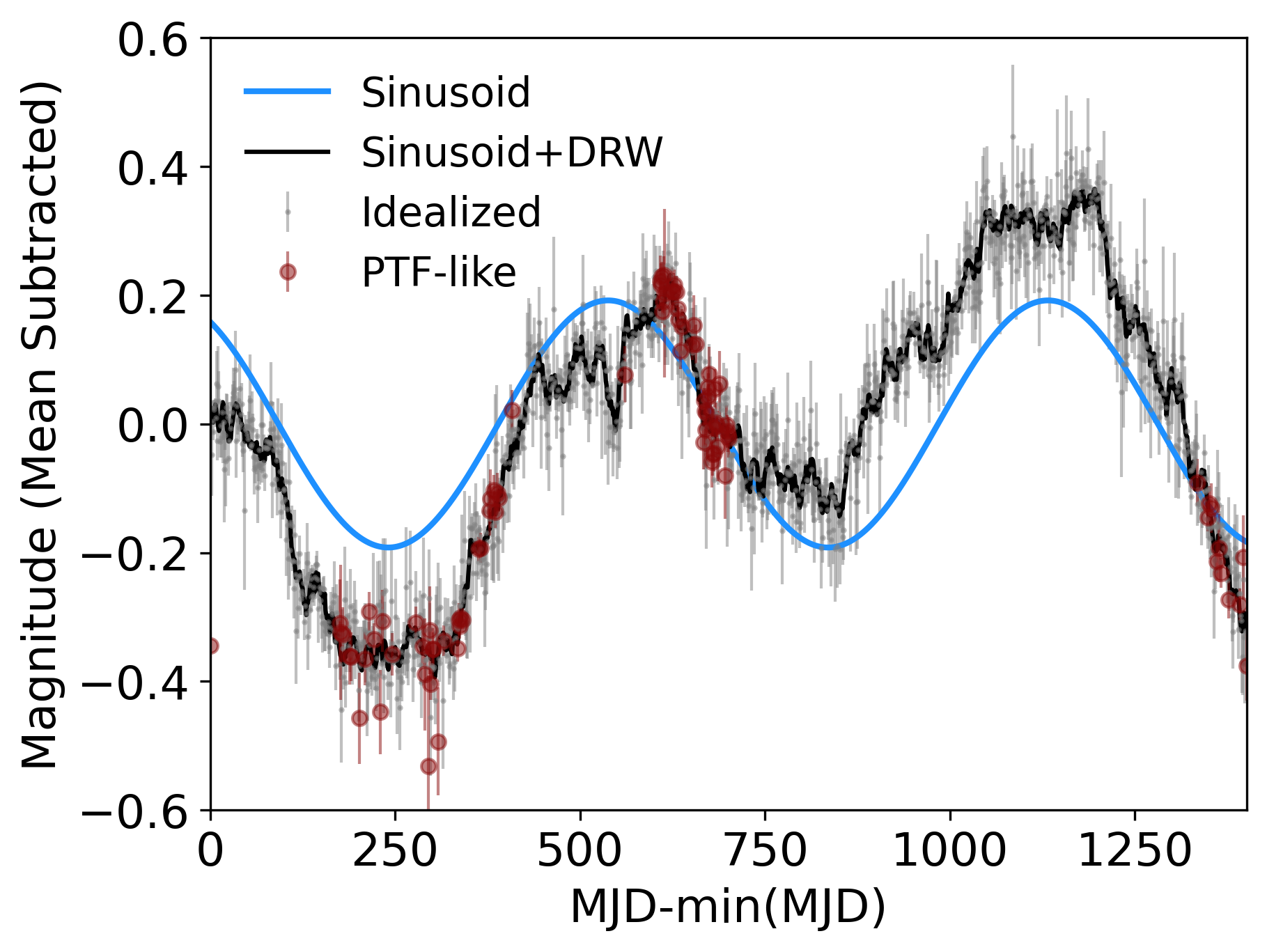}
    \includegraphics[width=0.45\textwidth]{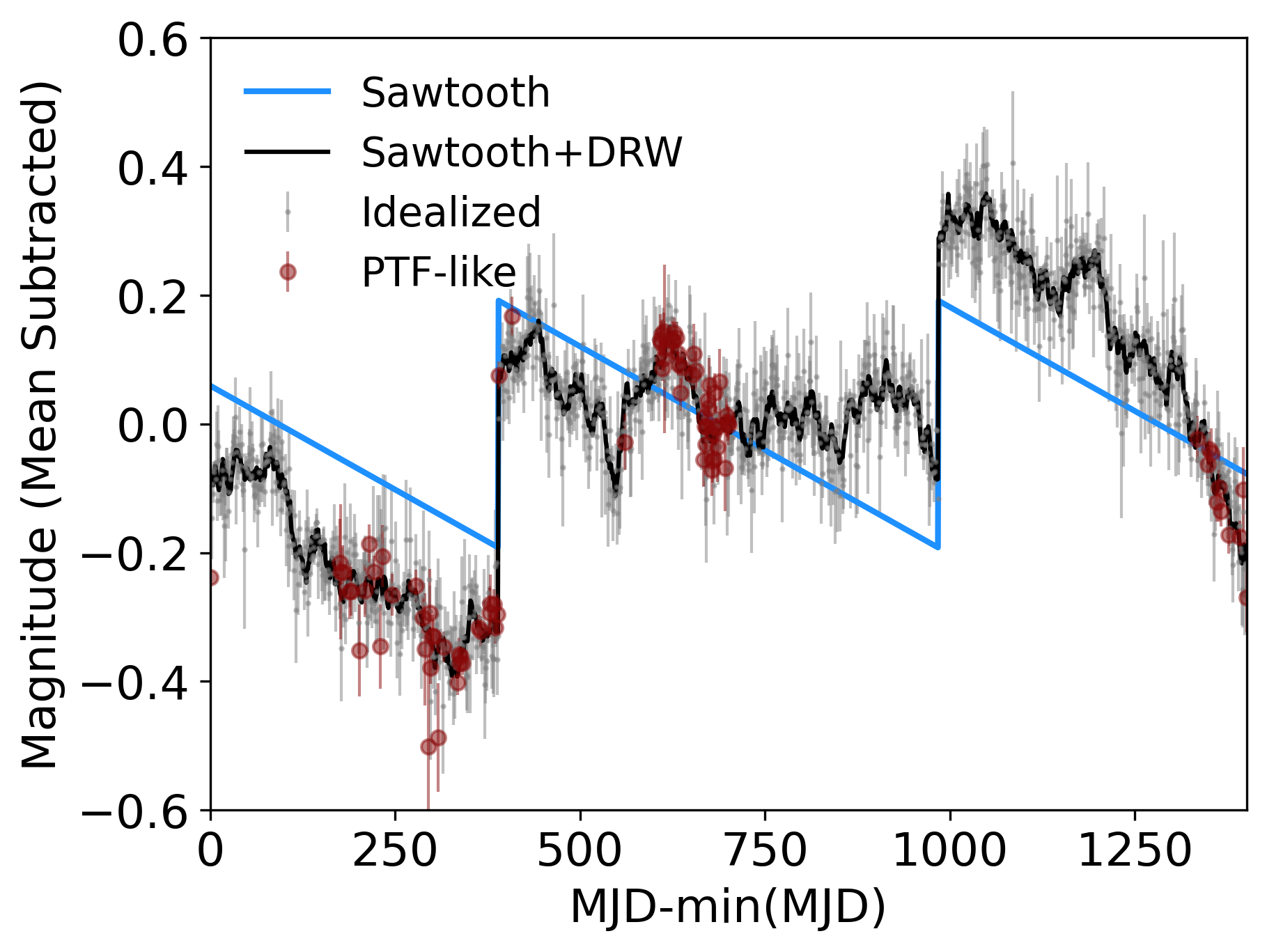}
    \caption{Examples of the mock periodic lightcurves we generate to test the detectability of such signals with the LSP, with a sinusoid injected on the left panel and a sawtooth signal injected on the right panel. The noiseless periodic signals are shown with solid blue lines. On top of those we add DRW noise (the same realization in both panels), shown with the black continuous curve. Idealised lightcurves sampled daily with photometric errors are shown with gray points and the respective error bars, while the PTF-like lightcurves shown in red are obtained by downsampling the black curve to the observed times and adding PTF-like photometric errors. The simulations are sampled at the observed times of quasar SDSS J085234.09+330100.7 with an injected period of 595 days,  amplitude 0.19 mag and DRW variability with parameters $\sigma=0.27$ mag and $\tau=1,246$ days.}
    \label{Fig:Lightcurve_examples}
\end{figure*}

We remind the reader that PTF was one of the early time-domain surveys (2009-2012), the precursor to the ongoing ZTF. It covered around 3,000 deg² with a 5$\sigma$ limiting magnitude of approximately 20.5 in the $R$-band.
Its observing strategy was designed to capture transient phenomena 
and thus 
the PTF lighcurves are characterised by epochs of very dense temporal sampling (e.g., when an interesting transient was detected in the field) followed by large gaps.
For a detailed discussion about the survey and the scientific goals of PTF see \citet{Law2009} and \citet{Rau2009}.

In this paper, we start from the sample of 35,383 quasars analyzed in \citet{Charisi2016}. This sample was extracted from the Half Million Quasars catalog, and then focusing only on the  spectroscopically confirmed quasars with relatively good lightcurves, i.e. lightcurves that have at least 50 nights of observations.
Of those, we select a subset of 12,400 that have virial SMBH mass estimates measured from their broad emission lines with spectra from the Sloan Digital Sky Survey (SDSS) Data Release 7 \citep{SDSS2009}. This selection is necessary for the estimation of the DRW parameters, as explained below (\S~\ref{sec:drw_sims}).

The PTF lightcurves for this subset of quasars  show a variety in terms of baselines and total number of observations (e.g., see Figure~1 in \citealt{Charisi2016}). 
On average, the baseline for these lightcurves is $\sim$3.2 years with a mean cadence of $\sim$12 days, a mean magnitude of 19.1 and mean photometric errors of $\sim$0.05.

\subsection{Mock lightcurves}
\label{subsec:LCs}
The lightcurves we simulate are composed of realistic quasar noise described by a DRW model (\S~\ref{sec:drw_sims}), \footnote{The DRW model, albeit successful, does not capture the full range of quasar variability. However, it provides a valuable benchmark for the effect of the red noise. In \S~\ref{sec:discuss}, we discuss some limitations of this choice of quasar variability.} periodic signals of two types to emulate two broad categories of binary behavior (\S~\ref{subsec:periodic_signals}) and photometric errors. For each of the 12,400 observed quasars in the sample, we generate a set of six mock lightcurves exploring two periodic signals: (a) sinusoidal periodicity and (b) sawtooth periodicity, and three observing scenarios: (1) idealised lightcurves and (2) PTF-like lightcuves, and (3) LSST-like lightcurves. 
For each set of six simulations, we inject the exact same parameters for the periodic signal (same period, amplitude and phase) and the same noise (i.e. the same DRW parameters $\sigma$ and $\tau$ and even the same noise realization).

For the idealised lightcurves, we simulate a time series with a baseline equal to the respective PTF lightcurve and photometric errors drawn from a Gaussian distribution with zero mean and standard deviation equal to the mean photometric error of the specific PTF lightcurve, but we adopt a daily cadence. Ground-based time-domain surveys cannot achieve such high-quality lightcurves, because uneven sampling and gaps are inevitable (e.g., missing observations due to weather or proximity of the source to the sun, etc). However, this set of idealised simulations (and the comparison with more realistic PTF-like lightcurves) allows us to isolate the effect the data quality has in detecting various periodic signals from other factors like the intrinsic quasar noise.

For the PTF-like lightcurves, we take the continuous time series (which include periodicity and DRW noise, but not photometric errors) from the above step and downsample them at the observed times of the respective PTF source. In this case, we add photometric errors drawn from a Gaussian distribution with zero mean and standard deviation equal to the photometric uncertainty of each point in the PTF lightcurve. 

Finally, we explore the periodicity recovery in longer lightcurves, which resemble the best-sampled single-band LSST lightcurves (e.g., what we expect for lightcurves in the $r$ or $i$-bands). 
We emphasize that LSST, with its unpredented data quality and a sample of quasars exceeding 20 millions, it will be a key driver in the discovery of SMBH binary candidates. 
For those, we set the baseline to 10 years, equal to the nominal duration of the survey. We choose a 5-day cadence, as expected for the Wide-Fast-Deep mode of LSST, but instead of evenly sampled data, we perturb the regular temporal sampling with a Gaussian of zero mean and standard deviation of 1 day. This choice allows us to more closely match the observing strategy of LSST, while avoiding strong aliasing effects from very regular sampling. We also exclude 4 months of observations every year to mimic the inevitable annual gaps. We add photometric errors drawing from a Gaussian distribution with zero mean and standard deviation of 0.01, which corresponds to the level of photometric uncertainty for sources of mean magnitude of 21 \citep{2019ApJ...873..111I}. 
Finally, we inject the same periodic signals and the same DRW realizations (creating them with the same random seed) as in the PTF lightcurves. We note that since the baselines are longer in the LSST-like lightcurves, we could explore longer periods. By simulating the same periods as in the shorter PTF lightcurves, we can explore the effect of observing more cycles within the data.

Therefore, for each observed lightcurve, we generate six mock lightcurves with the same injected periodicity and noise properties.
We analyze each of those lightcurves independently with a periodicity detection pipeline similar to the one presented in \cite{Charisi2016} (\S~\ref{2.3periodicity detection}).
Figure~\ref{Fig:Lightcurve_examples} shows such a set of four simulations idealised and PTF-like, and the components that produce the final lightcurve, with sinusoidal periodicity on the left panel and sawtooth periodicity on the right. With blue lines we demontrate the periodic signals and with black lines the aggregate time series after the DRW noise is included. Finally, the grey points with the error bars represent the idealised lightcurves after the photometric errors have been added, while the red points with the respective error bars show the PTF-like lightcurves produced from downsampling the black curve at the PTF timestamps and subsequently adding Gaussian errors from the PTF measurements.

\subsubsection{Quasar variability}
\label{sec:drw_sims}
In order to model realistic quasar lightcurves, we need to incorporate the intrinsic stochastic variability of quasars. 
The optical variability of quasars is successfully described by a DRW process \citep{Kelly2009,MacLeod+2010, Koz2010, 2021ApJ...907...96S, Burke_2021}. 
We emphasize that, albeit successful and widely used in previous periodicity searches, this model is likely an incomplete description of quasar variability (e.g., see \S~\ref{sec:discuss}). 
The choice of quasar noise model can affect our results and thus future studies should consider a more flexible description of the underlying variability (such as higher-order Continuous Auto-regressive Mean Average--CARMA--models; \citealt{2014ApJ...788...33K}), but the DRW model provides a useful baseline for red noise.

The power spectral density (PSD) of the DRW is frequency-dependent (red noise) at high frequencies, and becomes flat (white noise) at low frequencies. The PSD as a function of frequency $f$ can be described as follows: 
\begin{equation}
    PSD(f) = \frac{4\sigma^2\tau}{(1+4\pi^2f^2\tau^2)},
    \label{eq:PSD}
\end{equation}
where $\sigma^2$ is the variance of the lightcurve and $\tau$ is the damping timescale, roughly corresponding to the autocorrelation length of the lightcurve or to the (inverse of) the break frequency in the PSD. 

For each noise simulation, we first need to determine a set of DRW parameters, $\sigma$ and $\tau$. In \citet{Charisi2016}, the best-fit $\sigma$ and $\tau$ were determined for each lightcurve by maximizing the DRW likelihood using \texttt{Javelin}, a public software to sample the likelihood with Markov Chain Monte Carlo methods \citep{Javelin}.
Since these calculated $\sigma$ and $\tau$ values are no longer accessible, and that method resulted in biased estimates of the $\tau$ parameter (see Figure~8 and Section 4.1 in \citealt{Charisi2016}), we adopt a simpler alternative method to estimate these parameters. 

\begin{figure}
    \centering
    \includegraphics[width=0.45\textwidth]{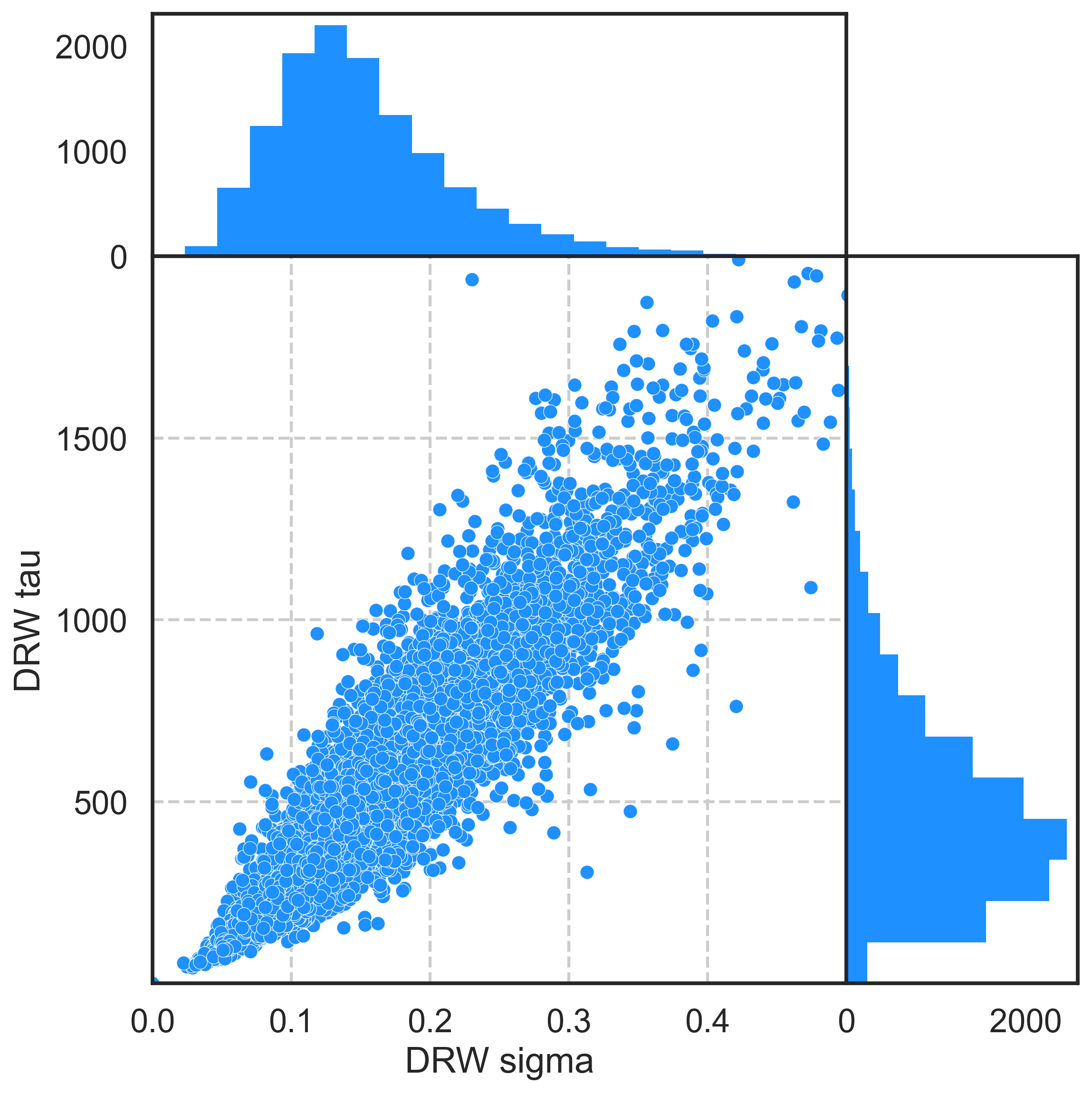}
    \caption{Values of the DRW parameters  $\sigma$ versus $\tau$ estimated based on their SMBH mass and $i$-band magnitude \citep{MacLeod+2010}
    for the 12,400 quasars in our sample. The marginalised distributions of the two quantities are also shown.}
    \label{Fig:sigma_tau}
\end{figure}

In particular, we use empirical scaling relationships from \citet{MacLeod+2010}, yielding $\sigma$ and $\tau$ from the properties of each individual quasar, as follows: 
\begin{align}
\log (\tau_{\mathrm{\rm RF}}) &= 2.4 + 0.17 \log \left( \frac{\lambda_{\mathrm{\rm RF}}}{4000\, \text{\AA}} \right) + 0.03 (M_i + 23) \notag \\
&\quad + 0.21 \log \left( \frac{M_{\mathrm{BH}}}{10^9 M_{\odot}} \right)
\label{tau}
\end{align}

\begin{align}
\log \left(\sqrt{2\sigma}\right) &= -0.51 - 0.479 \log \left( \frac{\lambda_{\mathrm{\rm RF}}}{4000\, \text{\AA}} \right) + 0.131 (M_i + 23) \notag \\
&\quad + 0.18 \log \left( \frac{M_{\mathrm{BH}}}{10^9 M_{\odot}} \right)
\label{sigma}
\end{align}
\noindent
where $\tau_{\rm RF}$ represents the characteristic time-scale $\tau$ in the quasar's rest frame, $\lambda_{\rm RF}$ is the effective wavelength of the $R$~band ($\lambda_R$ = 6516.05$\overset{\circ}{A}$)\footnote{PTF  observed in the optical $g$ and $R$ bands, but we only analyze the $R$-band lightcurves, as in \citet{Charisi2016}, because the $g$-band lightcurves are typically very sparse.} in the rest frame of the source, M$_{i}$ is the absolute i-band magnitude, k-corrected to $z$=2, and $M_{BH}$ is the SMBH mass. Note that these relationships were used as priors when fitting the DRW parameters with \texttt{Javelin} in \citet{Charisi2016}.
For the above $\sigma$ and $\tau$
calculations, the black hole mass and the $i$-band luminosity of each quasar are required. For this reason, as 
explained in \S~\ref{sec:sample}, we only analyze the sub-set of 12,400 quasars that have SMBH mass estimates and absolute $i$-band magnitudes, using a catalog from SDSS Data Release 7 \citep{SDSS2009}. 
Figure~\ref{Fig:sigma_tau} illustrates the distribution of the calculated $\tau$ and $\sigma$ values using Eq. (\ref{tau}) and (\ref{sigma}), respectively for the 12,400 PTF quasars.

For the mock quasar lightcurves, 
we follow the procedure from \citet{Timmer1995}, as implemented in the \texttt{astroML} package \citep{astroMLText}, and generate evenly sampled lightcurves with a temporal spacing of $\Delta$t = 1 day. As mentioned, in \S~\ref{sec:sample}, the baseline of each lightcurve is set to match the baseline of the respective observed lighcurve. 
For the PTF-like lightcurves, we follow a similar procedure as in \citet{Charisi2016} to downsample the idealised lightcurves by removing data points to match the observed pattern. For the LSST-like lightcurves, we generate the same realization of noise (using the same random seed), initially sampled daily for 10 years and then downsampled at the expected observing pattern of LSST (e.g., see \S~\ref{subsec:LCs}).

\subsubsection{Binary signals}
\label{subsec:periodic_signals}
For the simulations of the null hypothesis, required to assess the statistical significance of the injected periodic signals (see \S~\ref{2.3periodicity detection}), the simulations include only DRW variability, as outlined in the steps above. To simulate binary signals, we then add periodic variability. We inject two types of periodic signals, sinusoidal and sawtooth, in the following form: 
\begin{equation}
    \textbf{s}_{\mathrm{sine}}(t) = A \cdot\sin\left(2\pi\cdot t/P +\phi_0 \right),
\end{equation}
and
\begin{equation}
    \textbf{s}_{\mathrm{sawtooth}}(t) = A\cdot {\mathrm{sawtooth}}\left( 2\pi\cdot t/P + \phi_{0}  \right),
\end{equation} 
where A is the amplitude in magnitude, $\phi_0$ is the phase, and P is the period. The sawtooth waveform is characterised by an asymmetric sharp rise and linear decrease.

For the injected periodic signals, we explore a broad but realistic parameter space. In particular, for the periodic signals we draw parameters  from uniform distributions, from the following range of values:\\
$\bullet$ Amplitude A: [0, 0.5)\\
$\bullet$ Period P: [0, $P_{\rm max}$]\\
$\bullet$ Phase $\phi_0$: [0, 2$\pi$).\\
Here $P_{\rm max}$ is the maximum period we can probe in each lightcurve, defined by its baseline so that at least 1.5 cycles of periodicity are observed as in \cite{Charisi2016}. That is,
 $P_{\rm max}=T_{\mathrm{data}}/1.5$, where 
$T_{\mathrm{data}}$ is the baseline of the respective PTF lightcurve.\footnote{Based on the requirement of observing 1.5 cycles within the data, in the LSST-like lightcurves, we could have explored longer periods. However, we chose to explore the same periods as in the PTF-like data to isolate the effect of observing more cycles.} Since we randomly draw the parameters for each lightcurve and the baselines vary, the high end of the period distribution is not as well-sampled.

\subsection{Periodicity detection}
\label{2.3periodicity detection}
\begin{figure}
    \includegraphics[width=0.5\textwidth]{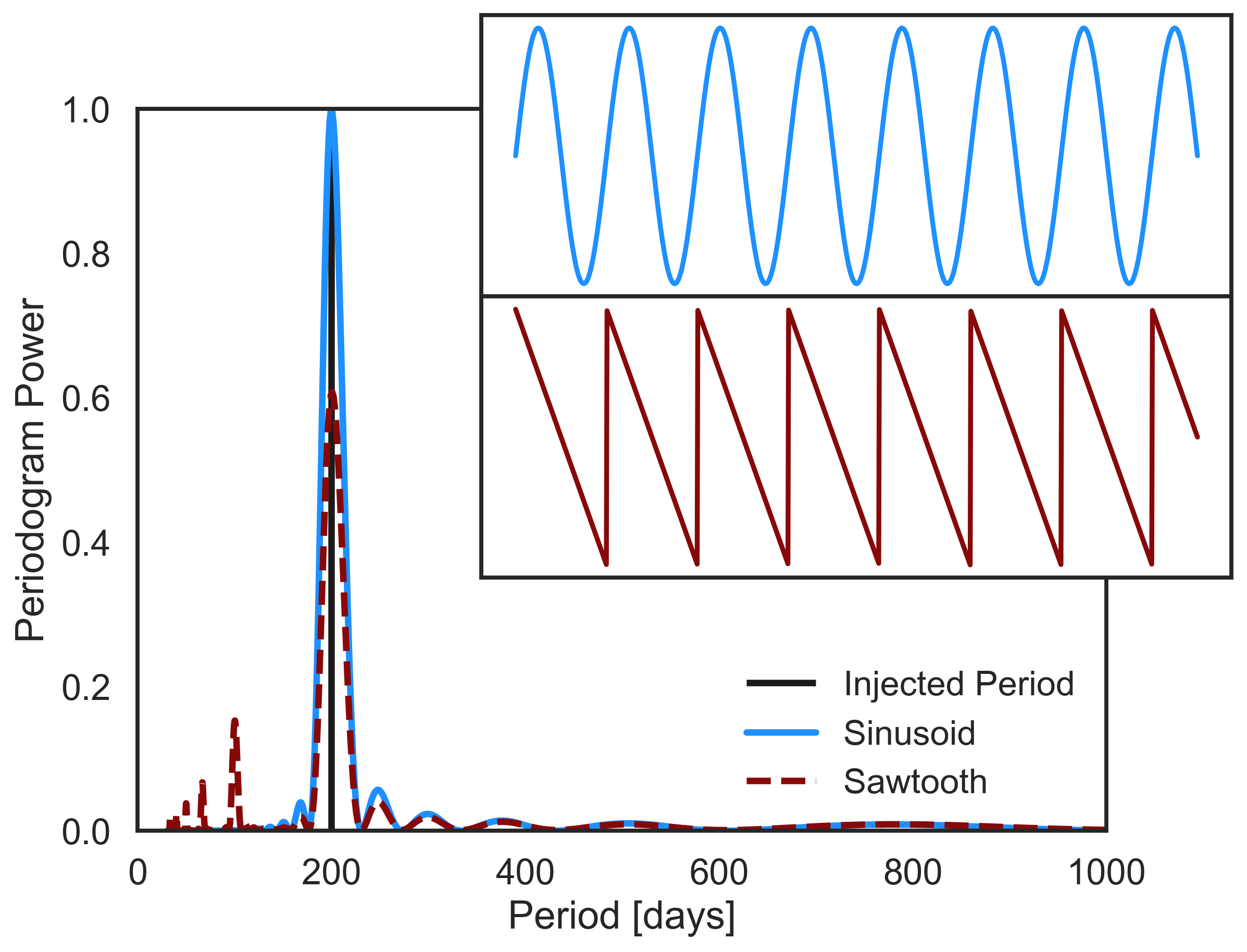}
    \caption{The Lomb-Scargle periodogram of a noiseless sinusoid (blue) and a sawtooth (red), both with a period of 200 days, and an amplitude of 0.4 shows peak power of 1 and 0.61, respectively.}
    \label{Fig:periodogram}
\end{figure}

Our goal is to assess whether the LSP is effective in detecting periodic signals that deviate from sinusoidal (e.g., sawtooth signals) in quasar data exploring several scenarios for lightcurve quality. For our analysis, we use the generalised LSP from the \texttt{astroML} package \citep{astroMLText}. 
We follow the same procedure as in \citet{Charisi2016} and calculate the LSP for 1,000 trial frequencies on a logarithmic grid with a minimum frequency of $f_{\mathrm{min}} = 1.5/T_{\mathrm{data}}$, 
and a maximum frequency of $f_{\rm max} = 1/T_{\mathrm{min}}$, with $T_{\mathrm{min}}$=30 days. 
In Figure~\ref{Fig:periodogram}, we illustrate how the LSP performs on two noiseless idealised lightcurves with sinusoidal and sawtooth shaped periodic signals, with amplitude of 0.4 mag and period of 200 days. We see that the peak of the periodogram is reduced by $\sim40\%$ for the sawtooth signal, while some secondary peaks at lower periods appear. These peaks are not from aliasing, because they are not present in the LSP of the sinusoidal signal but rather simply additional Fourier components of the sawtooth pattern.

Even though the false alarm probability of an LSP peak can be calculated analytically in the presence of white noise (e.g., see \citealt{Baluev_2008, Koen_2021}), these approximations are not applicable in the presence of other types of stochastic variability \citep{Suveges_2015, Vaughan2016}. Here, 
we assess the periodicity significance of the injected signals, again following \citet{Charisi2016}. We simulate DRW-only lightcurves representing the null-hypothesis, with $\sigma$ and $\tau$ chosen as described above, and compute the distribution of the periodogram peak heights in these mock lightcurves. Note that in a realistic scenario, the values of the DRW parameters would not be known, but our goal here is to determine the limitations from using a sub-optimal statistic (like the peak of the LSP), and so we assume we know the correct DRW parameters to simplify the analysis.
We compare the observed peaks of the injected sine+DRW and saw+DRW lightcurves against the observed peaks of DRW-only simulations.

To account for the fact that the DRW noise is frequency-dependent and thus the peaks in the periodogram do not arise with the same probability (or power) at every frequency, the test statistic needs to be compared with the null distribution in a small range of frequencies (i.e. we cannot observe a peak at low frequencies and compare with the null distribution at high frequencies and vice versa). For this, we divide our frequency grid into 25 different bins and for every bin, we calculate the false alarm probability of the maximum observed peak in that bin with the distribution of maximum peak heights of the simulated noise in the same bin. For each quasar, we simulate 100,000 DRW-only time series to account for the look-elsewhere effect, from the fact that the candidates were selected from a sample of 12,400 lightcurves and from using 25 frequency bins (see \citealt{Charisi2016} for more details).\footnote{In \cite{Charisi2016}, the false alarm probability was assessed with 250,000 simulations of DRW noise. However, since here we analyze only $\sim 1/3$ of that sample, we adjust the number of mock realizations to 100,000 to roughly reflect the smaller size analyzed.}
We consider the periodicity to be statistically significant if and only if a peak in any of the 25 frequency bins is higher than any of the peaks over the 100,000 DRW realizations in the same frequency bin. 

\section{Results}
\label{sec:results}
We assess the ability of the LS periodogram to detect sinusoidal and sawtooth periodicity in quasar data. For this, we simulate 12,400 mock lightcurves with idealised, PTF-like, and LSST-like observing patterns, injecting periodic signals of both types with a variety of parameters and realistic quasar noise. Then we assess the statistical significance of identified periodograms peaks with DRW simulations. We select as significant the ones that had periodogram peaks higher than the ones produced by 100,000 noise realizations.  
Below we report our findings.

\begin{figure*}
    \centering
    \includegraphics[width=\textwidth]{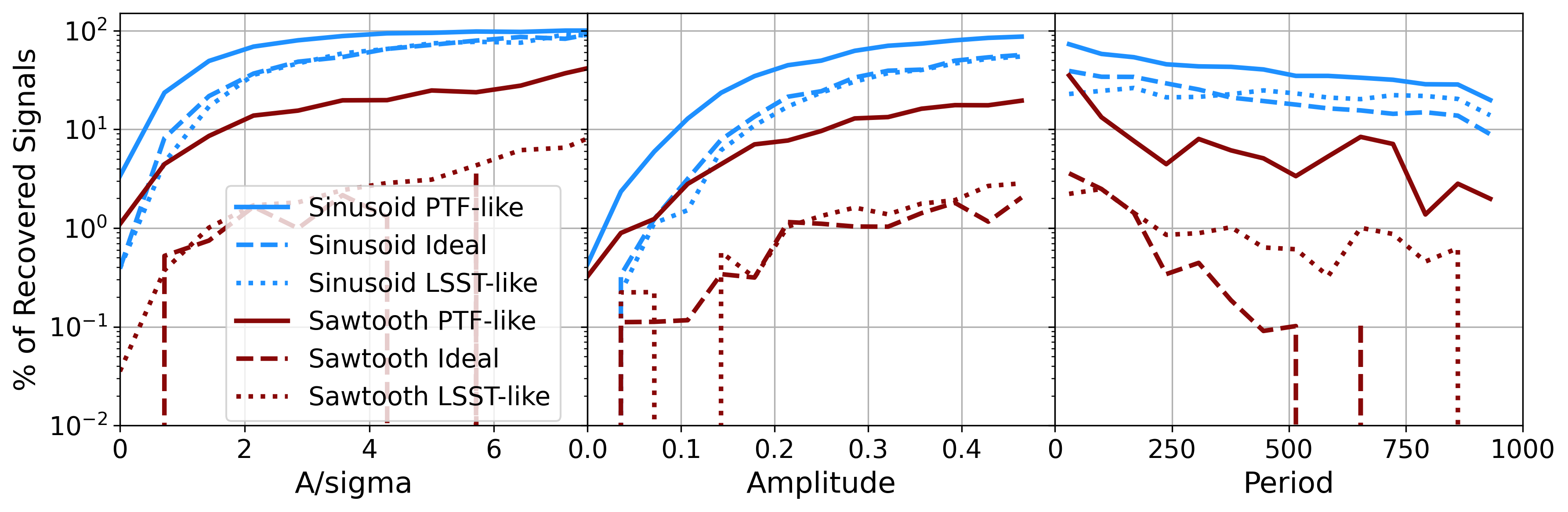}
    \caption{Recovery rates of periodic signals as a function of their properties. The left panel shows the dependence on the signal-to-noise ratio of the signal, quantified as the ratio of the amplitudes of the injected periodic signal versus the DRW noise ($A/\sigma$). The middle panel shows the dependence on the amplitude and the right panel on the period. The blue lines show the recovery rates for sinusoidal signals, while the red lines for sawtooth signals. Solid lines represent samples with PTF-like lightcurves, dashed lines are for idealised lightcurves, and dotted lines are for LSST-like simulations}.
    \label{Fig:Detection_Rates}
\end{figure*}

\subsection{Periodic signal recovery rates}
\label{subsec:detection-rate}

Among the 12,400 simulated sinusoidal signals, 3,020 (24.4\%)
are detected in the idealised lightcurves, 
5,540 (44.7\%) in the PTF-like simulations and 2,809 (22.6\%) in the LSST-like lightcurves. Interestingly, and somewhat counter-intuitively, the recovery rate of sinusoidal signals is higher in the PTF-like 
lightcurves. This is likely attributable to the underlying stochastic variability of quasars, as we discuss further in \S~\ref{sec:discuss} below.
We see that even though the LSP is one of the most prominent tools to detect sinusoidal periodicity in sparse lightcurves, it under-performs in quasar periodicity searches, missing $\sim$55\% of the sinusoidal signals in the PTF-like lightcurves and an even larger majority of $>$75\% in the idealised and LSST-like lightcurves. 

The reason for this poor performance has been identified by \cite{Robnik2024}, who recently demonstrated that the periodogram peak is a sub-optimal test statistic for this case. This is because its underlying assumption is white noise, while quasars have red noise variability. 
As a result, the best-fit Fourier coefficients, which are selected based on $\chi^2$ values that assumes the incorrect noise, tend to be biased. 
We emphasise that the stochastic variability can still be properly taken into account, e.g., by using red noise simulations to calculate the false alarm probability, as in \cite{Charisi2016}, but the test statistic itself is less sensitive in identifying these signals. For instance, using the Bayes factor (which explicitly incorporates the red noise) as the test statistic results in a higher true positive rate for the same false positive rate (see Figure~6 in \citealt{Robnik2024}). 
Similarly, using the Bayesian Information Criterion to select sinusoidal periodicity on top of DRW variability (i.e. including the red noise in the test statistic) \cite{Witt2022} has found higher overall recovery rates of $\sim$60\% in both CRTS-like and LSST-like simulated lightcurves.

The recovery rate of the sawtooth signals is significantly worse, with only 103 of the 12,400 (0.8\%) injections recovered in the idealised simulations, 1,150 (9.3\%) in the PTF-like lightcurves and 139 (1.1\%) in the LSST-like data.  These rates are summarised in Table~\ref{Tab:recovery_rates}, along with the results for sinusoidal signals. As with the sinusoidal signals, we detect more of the injected sawtooth signals in the PTF-like sample compared to the idealised and LSST-like case. The large reduction in the recovery rate between sinusoids and sawtooth signals is not surprising, given that a single peak in the LSP effectively uses a sinusoidal template and thus is optimised only for such signals (e.g., see Figure~\ref{Fig:periodogram}). Our results demonstrate that the LSP is very ineffective to find non-sinusoidal periodicity. It also implies
that if sawtooth signals were present in the PTF sample, it is highly unlikely that the previous analysis by \cite{Charisi2016} would have uncovered them. Although we do not explicitly demonstrate this here, the same conclusion very likely applies to the other previous periodicity searches listed in the Introduction, since most of those were also based on LSP. The exception is \citet{Graham2015}, where a combination of wavelets and the auto-correlation function (ACF) was employed. However, we suspect that their analysis would also have been insensitive to strongly non-sinusoidal periodicities, because of the chosen shape of wavelets (modulated plane waves) and the ACF (corresponding to a truncated Fourier series).

Finally, we examine whether the period of the detected signals is correctly identified. For this, we check whether the recovered signals have periods within 15\% of the simulated value. For the sinusoids we find that the vast majority, i.e. 98.8\% in the idealised lightcurves, 97.0\% in the PTF-like lightcurves and 99.7\% in the LSST-like lightcurves, have periods similar to the injected value. The fraction is slightly lower for the sawtooth signals, with 91.3\% in the idealised lightcurves, 79.6\% in the PTF-like lightcurves and 87.8\% in the LSST-like lightcurves. As seen in Figure~\ref{Fig:periodogram}, the periodogram of the sawtooth signals shows additional peaks beyond the injected period, which in combination with the red noise can lead to misidentification of the periods. In the PTF-like lightcruves, the uneven sampling can also contribute to this transfering of power between frequencies (known as aliasing). However, the misidentification of periods is overall relatively rare, especially for the sinusoids, for which the LSP is optimised.

\begin{table}
\centering
\begin{tabular}{|c|c|c|}
\hline
 & Sinusoid & Sawtooth \\
\hline
Idealized & $\sim$24\% & $\sim$1\% \\
\hline
PTF-like & $\sim$45\% & $\sim$9\% \\
\hline
LSST-like & $\sim$23\% & $\sim$1\%\\
\hline
\end{tabular}
\label{Tab:recovery_rates}
\caption{Recovery rate for sinusoidal and sawtooth periodic signals and for idealised, PTF-like, and LSST-like} lightcurves.
\end{table}

\subsection{Dependence on signal parameters}

We next explore how the recovery rates depend on the parameters of the injected periodic signals, in combination with the DRW noise properties. First, we quantify the signal strength with the ratio of the amplitude of the periodic signal over the variance of the DRW noise, $A/\sigma$, which provides a rough estimate of the signal-to-noise ratio (SNR). In the left panel of Figure~\ref{Fig:Detection_Rates}, we show the recovery rates as a function of their SNR, with blue lines for sinusoidal signals, and red for sawtooth signals. Solid lines represent the PTF-like simulations, while dashed lines indicate the idealised lightcurves, and dotted lines the LSST-like simulations.
We see that the recovery rate increases as the SNR increases, as expected, i.e. stronger signals are easier to detect compared to signals buried in the DRW noise. The trend is roughly monotonic for both types of signals and quality of lightcurves, with the exception of sawtooth signals in idealised lightcurves, where we observe some fluctuations at the relatively poorly sampled high SNR regime due to small number statistics.

Next, we consider the recovery rates in terms of the properties of the periodic signal alone. The middle panel of Figure~\ref{Fig:Detection_Rates} shows the recovery rates as a function of the injected amplitude of the periodic signal, with the same colors and line styles as above.
Again, we observe monotonically increasing rates for both types of periodic signals and for all three types of lightcurves, i.e. signals with higher amplitudes are significantly easier to detect compared to low-amplitude signals, as expected.
Even in the best case scenario, that is, very strong signals with $A=0.5$, the algorithm does not detect all the signals, reaching a maximum recovery rate of 80\%. This also demonstrates the limitations of the LSP to detect periodic signals in the presence of red noise. 

Finally, we explore the recovery rates as a function of the period. We observe decreasing trends with the period in all six cases we examined. 
In particular, for the sinusoids the decrease is monotonic, whereas for the sawtooth signals, we see a sharp decline and the signals become practically undetectable for periods longer than 250 days for the idealised lightcurves, with fluctuations at the level of a few percent
for the PTF-like and the LSST-like simulations.
This trend is also expected, since a shorter period signal allow for more cycles to be observed within the available baseline, thus resulting in stronger periodogram peaks and higher probability of detection, whereas longer-period signals, with only 2-3 observed cycles are very challenging to detect. In this regime, the periodogram peaks are also relatively broad, resulting in significant uncertainty in the period estimation.
Additionally, the DRW noise amplitude is lower at higher frequencies (see Eq.~\ref{eq:PSD}).
This limitation is significant because in a population of binaries we expect many more long-period binaries than short-period ones. Therefore, the samples of candidates in this very interesting part of the parameter space are likely highly incomplete. Again, even in the best-case scenario, of periods of 30 days, which is the minimum searched period in this study, the recovery rate does not reach 100\%, underscoring the challenges of detecting real signals. Similar trends for the recovery rates with respect to the amplitude \citep{Witt2022,2024ApJ...965...34D} and the period of the signal \citep{Witt2022} have been seen before.

\section{Discussion}
\label{sec:discuss}

Our study focuses on the detection of non-sinusoidal periodic signals in quasar lightcurves using the LSP under different scenarios for lightcurve quality. Such signals are expected to be common in binary lightcurves, but may have been missed in previous searches, which focused primarily on sinusoidal periodicity. To quantify the expected detection rates, we injected sinusoidal and sawtooth-shaped periodic signals in realistic  PTF-like, idealised and LSST-like lightcurves, which are expected to be crucial in the near-future. Using a statistical analysis similar to the one employed in \cite{Charisi2016}, which is based on the LSP, we compute the recovery rate for both signals and for all three types of lightcurve data quality.

Our main conclusion is that the LSP struggles to detect the injected periodicity, detecting a maximum of $\sim$45\% of the sinusoidal signals and only up to $\sim$9\% of the sawtooth signals in the PTF-like lightcurves. We conclude that previous searches, including the one in \cite{Charisi2016}, must have missed a significant fraction of periodic signals. This is the case even for sinusoids, but the searches likely missed the vast majority of binaries if they produce sawtooth-shaped periodicity, as hydrodynamical simulations suggest. 
Caution is required in generalizing this conclusion, because the recovery rates depend on the specific periodicity detection method, precise pulse shape, as well as on the data quality. However, all previous searches used Fourier algorithms which, like the LSP, are optimised for sinusoidal signals. 

Our analysis also returned a surprising (to us at least) finding; the LSP performs better in the PTF-like sample compared to the idealised or LSST-like ones, both for sinusoidal and sawtooth periodicity, although with significantly lower overall recovery rates for the latter. One would expect that the recovery rate would be higher in lightcurves with better quality and longer baselines (for the LSST-like simulations). However, as illustrated in Figure~\ref{Fig:Lightcurve_examples}, the inclusion of DRW noise results in lightcurves that significantly deviate from the simple periodic behavior. The LSP, which performs a weighted least-squares fit for sinusoid amplitudes at each searched frequency, will struggle to produce a good fit, resulting in a lower/less significant peak. However, if one considers the sparser PTF-like lightcurve, it is easier to construct a sinusoidal curve (e.g., by adjusting the mean or the amplitude) that passes through the red points. We conclude that in the PTF-like lightcurves, the sparse sampling can in many cases be advantageous, i.e. if by chance the observed epochs sample close to the maxima and minima of the lightcurve. On the other hand, we caution that here we considered only real signals and did not consider the case of false detections, which presents a major challenge in quasar periodicity searches. This is likely exacerbated by the lower quality of lightcurves (e.g., see the comparison of CRTS-like versus LSST-like simulations in \citealt{Witt2022}).

Moreover, our study showed that for LSST-like lightcurves, the LSP detects $<$25\% of the sinusoidal signals, and $\sim$1\% of the sawtooth periodicity. This is a significant finding for periodicity searches in the upcoming LSST data clearly demonstrating that the development of tailored periodicity tools that explicitly account for red noise variability is necessary.
As we approach the era of large time-domain surveys, with the Rubin Observatory \citep{2019ApJ...873..111I} set to begin operations in just a few months, it is evident that new methods for detecting periodicity in quasars are required. LSST will provide an unprecedented dataset with $\sim 20$ to 100 million quasars in which periodic signals are expected to be abundant
\citep{2021MNRAS.506.2408X,2021MNRAS.508.2524K,2024A&A...691A.250C}. Our ability to detect those relies on developing methods that have a high true positive rate (unlike the LSP), while simultaneously reducing the false positive rate as much as possible (e.g., \citealt{Robnik2024}). 
We demostranted that even in LSST-like lightcurves, which have improved quality and longer baselines compared to PTF, the recovery rates are low. Therefore, more sophisticated periodicity search algorithms, specifically targeting non-sinusoidal periodicity, e.g., methods that employ matched-filtering or search for periodicity in phase-folded lightcurves may offer significant advantages compared to the LSP (e.g., see \citealt{2013MNRAS.434.3423G} for a comparison of methods). Here we demonstrated that one of the main tools, the LSP, performs very poorly, especially with sawtooth periodicity.  %Beyond new techniques, we also expect that the improved data quality of LSST will be beneficial. The PTF-like lightcurves used in this study are very sparse, with highly irregular sampling. The ongoing ZTF survey already delivers higher quality data, while LSST will further improve the data quality with lightcurves characterised by superior photometric precision, longer baselines, and a superb nominal cadence of 3-5 days.
%The large sample, combined with the superior lightcurve quality is expected to enhance the reliability and sensitivity of periodicity searches in quasar lightcurves \citep{Witt2022}, potentially also uncovering a broader range of periodic behaviors.

As we already mentioned, the stochastic nature of quasar variability is a major limitation. In this study, we used the DRW model to simulate the underlying noise in quasars. This model provides a successful description of optical variability \citep{Kelly2009,MacLeod+2010, Koz2010, 2021ApJ...907...96S, Burke_2021}, but recent work has found deviations from this simple model \citep{2015MNRAS.451.4328K,2018ApJ...857..141S,Robnik2024}. A higher order stochastic model, such as the damped harmonic oscillator \citep{2019PASP..131f3001M, 2022ApJ...936..132Y} or higher order CARMA models \citep{2014ApJ...788...33K} may provide more flexibility and a better description of quasar variability.  Improvements in our understanding of quasar variability will need to be incorporated in future periodicity searches to increase their sensitivity and to reduce the false detections. It is worth noting that our assumptions about the underlying noise may impact the reported recovery rates of periodic signals, but the overall trends will likely remain unaffected. We plan to explore additional noise models in a future study.

In addition, throughout our analysis, we adopted a simple sawtooth-shaped periodicity. This cartoon model is sufficient for the purposes of our exercise, and illustrates the large impact of non-sinusoidal pulse shapes.  However, this shape was merely inspired by the results of hydrodynamical simulations, and the actual shapes of lightcurves produced by binary SMBHs will differ from this shape -- indeed it is poorly understood and will likely vary from source to source.  Developing better templates of the expected light-curve shapes \citep[e.g., see][]{binlite} can be used to better characterise the completeness and purity of binary candidate samples, and will likely allow improvements in recovery rates.

Finally, another possible limitation in the search for quasar periodicity is unmodeled photometric effects that can introduce spurious signals at the annual or lunal cycle. Such artificial periodicities, leftover from the data reduction process, can be convolved with real periodic signals from SMBHBs impacting their detection (e.g., introducing false positives, moving power to other frequencies hindering the detection or leading to miss-identification of the real periodicity, etc). Previous quasar periodicity searches have not explored this effect, but discarded periods very close to 1 year and have completely avoided periods $<$1 month. However, it is important to understand and mitigate such effects. In a future study, we plan to inject periodic signals in observed quasar data and explore this in several photometric surveys, like CRTS, PTF, ZTF, PanSTARRS, etc, including also some of the AGN that have long-term monitoring (e.g., \citealt{2002ApJ...581..197P, 2021arXiv210600691K, 2023MNRAS.520.1807C}).

\section{Conclusions}
\label{sec:conclusion}
In this study, we explore the efficacy of the LS periodogram (one of the most popular tools to search for periodic signals) in detecting non-sinusoidal periodicity in quasars. Periodic signals that significantly deviate from sinusoids, like sawtooth periodicity, are often seen in hydrodynamical simulations of SMBHBs and may be common among quasars, but would likely have been missed in observational searches to date. 

We start from the sample of 35,383 quasar lightcurves from PTF, previously analyzed in \cite{Charisi2016}, and select a sub-set of 12,400 for which the SMBH masses are measured. This allows us to estimate the DRW parameters from scaling relations \citep{MacLeod+2010}. 
Since LSST will soon provide a large dataset of high-quality quasar lightcurves for SMBH binary searches, we also consider LSST-like lightcurves.
We 
simulate quasar lightcurves with DRW noise, but note that this is an approximation to quasar noise. We inject two types of periodicity: (a) sinusoidal signals and (b) sawtooth signals. We explore three types of lighcurves to examine the limitations of LSP under different survey conditions: (1) PTF-like lightcurves, with properties (photometric errors, sampling, baselines) resembling the observational data, (2) idealised lightcurves with daily cadence, baselines equal to the respective PTF lightcurves and Gaussian photometric errors with standard deviations equal to the mean photometric error of each source, and (3) LSST-like lightcurves with a baseline of 10 years, semi-regular cadence of 5 days, annual gaps of 4 months and photometric errors, as expected for a source of magnitude $\sim$21. For the detection of periodicity, we follow the statistical analysis of \cite{Charisi2016}, simulating DRW lightcurves to assess the false alarm probability of identified signals.

Our main findings are summarised as follows:\\
\textbullet\ 
The periodogram peaks are reduced by $\sim$40$\%$, when comparing noiseless sawtooth versus sinusoidal signals of the same period and amplitude injected in idealised lightcurves.\\
\textbullet\ 
Of the sinusoidal signals, $\sim$45\% are recovered in PTF-like lightcurves, $\sim$25\% in the idealised lightcurves, and $\sim$23\% in the LSST-like simulations. The stochastic variability of quasars prevents us from identifying the majority of the injected sinusoids.\\
\textbullet\ 
The fraction of recovered signals is significantly worse for sawtooth periodicity, with only $\sim$9\% of the signals detected in the PTF-like lightcurves, $\sim$1\% for the idealised simulations, and $\sim$1\% for the LSST-like simulations. This suggests that the LSP misses the vast majority of binaries, and is not an appropriate tool to search for this type of periodicity.\\
\textbullet\ 
The recovered periods are within 15\% of the injected values for the vast majority of detected signals, with $\sim$99\% ($\sim$91\%) for sinusoidal (sawtooth signals) in the idealized lightcurves, $\sim$97\% ($\sim$80\%) in the PTF-like data, and $\sim$100\% ($\sim$88\%) in the LSST-like lightcurves.\\
\textbullet\ The recovery rates depend on the signal and DRW noise properties as expected. They increase monotonically with the SNR (quantified as $A/\sigma$), and the amplitude of the periodic signals, and thus strong signals are relatively easy to detect, even in the presence of stochastic variability.\\
\textbullet\ The recovery rates decrease as a function of the injected period, which means that the rarer short-period binaries are easier to detect compared to the more abundant long-period ones. For sawtooth signals, binaries with periods $>$250 days become almost undetectable (with a detection rate $>$1\% for idealized and LSST-like lightcurves), primarily due to the DRW noise, which is more significant at these low frequencies.\\~

In this study, we found that non-sinusoidal periodic signals, particularly those with a sawtooth-shaped periodicity predicted in hydrodynamical simulations, can be easily missed in periodicity searches that rely on the LSP. The LSP struggles even with the sinusoidal signals, detecting less than 45\% in all three types of lightcurves. This is likely due to the stochastic variability of quasars which makes the peak of the LSP a sub-optimal test statistic \citep{Robnik2024}. Based on the above, we can conclude that previous searches would struggle to detect sawtooth signals, even if they were present in the samples. They would have missed a significant fraction of binaries even in the limit that they produced pure sinusoidal signals. This, in combination with the fact that the stochastic variability can introduce false detections, underscores the challenges in identifying periodicity in quasars. Such challenges must be addressed with improved periodicity detection methods in future searches, in order to take full advantage of the upcoming time-domain data.

\label{sec:conclude}
\section*{Acknowledgements}
MC acknowledges support by the European Union (ERC, MMMonsters,
101117624).
This work was also supported in part by NASA grants 80NSSC24K0440 and 80NSSC22K0822.
This research used resources of the Center for
Institutional Research Computing at Washington State University.

\bibliographystyle{aasjournal}
\bibliography{mma}

\begin{thebibliography}{}
\expandafter\ifx\csname natexlab\endcsname\relax\def\natexlab#1{#1}\fi
\providecommand{\url}[1]{\href{#1}{#1}}
\providecommand{\dodoi}[1]{doi:~\href{http://doi.org/#1}{\nolinkurl{#1}}}
\providecommand{\doeprint}[1]{\href{http://ascl.net/#1}{\nolinkurl{http://ascl.net/#1}}}
\providecommand{\doarXiv}[1]{\href{https://arxiv.org/abs/#1}{\nolinkurl{https://arxiv.org/abs/#1}}}

\bibitem[{{Abazajian} {et~al.}(2009){Abazajian}, {Adelman-McCarthy}, {Ag{\"u}eros}, {Allam}, {Allende Prieto}, {An}, {Anderson}, {Anderson}, {Annis}, {Bahcall}, {Bailer-Jones}, {Barentine}, {Bassett}, {Becker}, {Beers}, {Bell}, {Belokurov}, {Berlind}, {Berman}, {Bernardi}, {Bickerton}, {Bizyaev}, {Blakeslee}, {Blanton}, {Bochanski}, {Boroski}, {Brewington}, {Brinchmann}, {Brinkmann}, {Brunner}, {Budav{\'a}ri}, {Carey}, {Carliles}, {Carr}, {Castander}, {Cinabro}, {Connolly}, {Csabai}, {Cunha}, {Czarapata}, {Davenport}, {de Haas}, {Dilday}, {Doi}, {Eisenstein}, {Evans}, {Evans}, {Fan}, {Friedman}, {Frieman}, {Fukugita}, {G{\"a}nsicke}, {Gates}, {Gillespie}, {Gilmore}, {Gonzalez}, {Gonzalez}, {Grebel}, {Gunn}, {Gy{\"o}ry}, {Hall}, {Harding}, {Harris}, {Harvanek}, {Hawley}, {Hayes}, {Heckman}, {Hendry}, {Hennessy}, {Hindsley}, {Hoblitt}, {Hogan}, {Hogg}, {Holtzman}, {Hyde}, {Ichikawa}, {Ichikawa}, {Im}, {Ivezi{\'c}}, {Jester}, {Jiang}, {Johnson}, {Jorgensen}, {Juri{\'c}}, {Kent}, {Kessler}, {Kleinman}, {Knapp},
  {Konishi}, {Kron}, {Krzesinski}, {Kuropatkin}, {Lampeitl}, {Lebedeva}, {Lee}, {Lee}, {French Leger}, {L{\'e}pine}, {Li}, {Lima}, {Lin}, {Long}, {Loomis}, {Loveday}, {Lupton}, {Magnier}, {Malanushenko}, {Malanushenko}, {Mandelbaum}, {Margon}, {Marriner}, {Mart{\'\i}nez-Delgado}, {Matsubara}, {McGehee}, {McKay}, {Meiksin}, {Morrison}, {Mullally}, {Munn}, {Murphy}, {Nash}, {Nebot}, {Neilsen}, {Newberg}, {Newman}, {Nichol}, {Nicinski}, {Nieto-Santisteban}, {Nitta}, {Okamura}, {Oravetz}, {Ostriker}, {Owen}, {Padmanabhan}, {Pan}, {Park}, {Pauls}, {Peoples}, {Percival}, {Pier}, {Pope}, {Pourbaix}, {Price}, {Purger}, {Quinn}, {Raddick}, {Re Fiorentin}, {Richards}, {Richmond}, {Riess}, {Rix}, {Rockosi}, {Sako}, {Schlegel}, {Schneider}, {Scholz}, {Schreiber}, {Schwope}, {Seljak}, {Sesar}, {Sheldon}, {Shimasaku}, {Sibley}, {Simmons}, {Sivarani}, {Allyn Smith}, {Smith}, {Smol{\v{c}}i{\'c}}, {Snedden}, {Stebbins}, {Steinmetz}, {Stoughton}, {Strauss}, {SubbaRao}, {Suto}, {Szalay}, {Szapudi}, {Szkody}, {Tanaka},
  {Tegmark}, {Teodoro}, {Thakar}, {Tremonti}, {Tucker}, {Uomoto}, {Vanden Berk}, {Vandenberg}, {Vidrih}, {Vogeley}, {Voges}, {Vogt}, {Wadadekar}, {Watters}, {Weinberg}, {West}, {White}, {Wilhite}, {Wonders}, {Yanny}, {Yocum}, {York}, {Zehavi}, {Zibetti}, \& {Zucker}}]{SDSS2009}
{Abazajian}, K.~N., {Adelman-McCarthy}, J.~K., {Ag{\"u}eros}, M.~A., {et~al.} 2009, \apjs, 182, 543, \dodoi{10.1088/0067-0049/182/2/543}

\bibitem[{{Agazie} {et~al.}(2023{\natexlab{a}}){Agazie}, {Anumarlapudi}, {Archibald}, {Arzoumanian}, {Baker}, {B{\'e}csy}, {Blecha}, {Brazier}, {Brook}, {Burke-Spolaor}, {Burnette}, {Case}, {Charisi}, {Chatterjee}, {Chatziioannou}, {Cheeseboro}, {Chen}, {Cohen}, {Cordes}, {Cornish}, {Crawford}, {Cromartie}, {Crowter}, {Cutler}, {Decesar}, {Degan}, {Demorest}, {Deng}, {Dolch}, {Drachler}, {Ellis}, {Ferrara}, {Fiore}, {Fonseca}, {Freedman}, {Garver-Daniels}, {Gentile}, {Gersbach}, {Glaser}, {Good}, {G{\"u}ltekin}, {Hazboun}, {Hourihane}, {Islo}, {Jennings}, {Johnson}, {Jones}, {Kaiser}, {Kaplan}, {Kelley}, {Kerr}, {Key}, {Klein}, {Laal}, {Lam}, {Lamb}, {Lazio}, {Lewandowska}, {Littenberg}, {Liu}, {Lommen}, {Lorimer}, {Luo}, {Lynch}, {Ma}, {Madison}, {Mattson}, {McEwen}, {McKee}, {McLaughlin}, {McMann}, {Meyers}, {Meyers}, {Mingarelli}, {Mitridate}, {Natarajan}, {Ng}, {Nice}, {Ocker}, {Olum}, {Pennucci}, {Perera}, {Petrov}, {Pol}, {Radovan}, {Ransom}, {Ray}, {Romano}, {Sardesai}, {Schmiedekamp}, {Schmiedekamp},
  {Schmitz}, {Schult}, {Shapiro-Albert}, {Siemens}, {Simon}, {Siwek}, {Stairs}, {Stinebring}, {Stovall}, {Sun}, {Susobhanan}, {Swiggum}, {Taylor}, {Taylor}, {Turner}, {Unal}, {Vallisneri}, {van Haasteren}, {Vigeland}, {Wahl}, {Wang}, {Witt}, {Young}, \& {Nanograv Collaboration}}]{2023ApJ...951L...8A}
{Agazie}, G., {Anumarlapudi}, A., {Archibald}, A.~M., {et~al.} 2023{\natexlab{a}}, \apjl, 951, L8, \dodoi{10.3847/2041-8213/acdac6}

\bibitem[{{Agazie} {et~al.}(2023{\natexlab{b}}){Agazie}, {Anumarlapudi}, {Archibald}, {Baker}, {B{\'e}csy}, {Blecha}, {Bonilla}, {Brazier}, {Brook}, {Burke-Spolaor}, {Burnette}, {Case}, {Casey-Clyde}, {Charisi}, {Chatterjee}, {Chatziioannou}, {Cheeseboro}, {Chen}, {Cohen}, {Cordes}, {Cornish}, {Crawford}, {Cromartie}, {Crowter}, {Cutler}, {D'Orazio}, {Decesar}, {Degan}, {Demorest}, {Deng}, {Dolch}, {Drachler}, {Ferrara}, {Fiore}, {Fonseca}, {Freedman}, {Gardiner}, {Garver-Daniels}, {Gentile}, {Gersbach}, {Glaser}, {Good}, {G{\"u}ltekin}, {Hazboun}, {Hourihane}, {Islo}, {Jennings}, {Johnson}, {Jones}, {Kaiser}, {Kaplan}, {Kelley}, {Kerr}, {Key}, {Laal}, {Lam}, {Lamb}, {Lazio}, {Lewandowska}, {Littenberg}, {Liu}, {Luo}, {Lynch}, {Ma}, {Madison}, {McEwen}, {McKee}, {McLaughlin}, {McMann}, {Meyers}, {Meyers}, {Mingarelli}, {Mitridate}, {Natarajan}, {Ng}, {Nice}, {Ocker}, {Olum}, {Pennucci}, {Perera}, {Petrov}, {Pol}, {Radovan}, {Ransom}, {Ray}, {Romano}, {Runnoe}, {Sardesai}, {Schmiedekamp}, {Schmiedekamp},
  {Schmitz}, {Schult}, {Shapiro-Albert}, {Siemens}, {Simon}, {Siwek}, {Stairs}, {Stinebring}, {Stovall}, {Sun}, {Susobhanan}, {Swiggum}, {Taylor}, {Taylor}, {Turner}, {Unal}, {Vallisneri}, {Vigeland}, {Wachter}, {Wahl}, {Wang}, {Witt}, {Wright}, {Young}, \& {Nanograv Collaboration}}]{2023ApJ...952L..37A}
---. 2023{\natexlab{b}}, \apjl, 952, L37, \dodoi{10.3847/2041-8213/ace18b}

\bibitem[{{Baluev}(2008)}]{Baluev_2008}
{Baluev}, R.~V. 2008, \mnras, 385, 1279, \dodoi{10.1111/j.1365-2966.2008.12689.x}

\bibitem[{{B{\'e}csy} {et~al.}(2022){B{\'e}csy}, {Cornish}, \& {Kelley}}]{2022arXiv220701607B}
{B{\'e}csy}, B., {Cornish}, N.~J., \& {Kelley}, L.~Z. 2022, arXiv e-prints, arXiv:2207.01607.
\newblock \doarXiv{2207.01607}

\bibitem[{{Begelman} {et~al.}(1980){Begelman}, {Blandford}, \& {Rees}}]{Begelman1980}
{Begelman}, M.~C., {Blandford}, R.~D., \& {Rees}, M.~J. 1980, \nat, 287, 307, \dodoi{10.1038/287307a0}

\bibitem[{{Bogdanovic} {et~al.}(2021){Bogdanovic}, {Miller}, \& {Blecha}}]{2021arXiv210903262B}
{Bogdanovic}, T., {Miller}, M.~C., \& {Blecha}, L. 2021, arXiv e-prints, arXiv:2109.03262.
\newblock \doarXiv{2109.03262}

\bibitem[{{Bon} {et~al.}(2016){Bon}, {Zucker}, {Netzer}, {Marziani}, {Bon}, {Jovanovi{\'c}}, {Shapovalova}, {Komossa}, {Gaskell}, {Popovi{\'c}}, {Britzen}, {Chavushyan}, {Burenkov}, {Sergeev}, {La Mura}, {Vald{\'e}s}, \& {Stalevski}}]{2016ApJS..225...29B}
{Bon}, E., {Zucker}, S., {Netzer}, H., {et~al.} 2016, \apjs, 225, 29, \dodoi{10.3847/0067-0049/225/2/29}

\bibitem[{{Burke} {et~al.}(2021){Burke}, {Shen}, {Blaes}, {Gammie}, {Horne}, {Jiang}, {Liu}, {McHardy}, {Morgan}, {Scaringi}, \& {Yang}}]{Burke_2021}
{Burke}, C.~J., {Shen}, Y., {Blaes}, O., {et~al.} 2021, Science, 373, 789, \dodoi{10.1126/science.abg9933}

\bibitem[{{Burke-Spolaor} {et~al.}(2019){Burke-Spolaor}, {Taylor}, {Charisi}, {Dolch}, {Hazboun}, {Holgado}, {Kelley}, {Lazio}, {Madison}, {McMann}, {Mingarelli}, {Rasskazov}, {Siemens}, {Simon}, \& {Smith}}]{Burke-Spolaor2019}
{Burke-Spolaor}, S., {Taylor}, S.~R., {Charisi}, M., {et~al.} 2019, \aapr, 27, 5, \dodoi{10.1007/s00159-019-0115-7}

\bibitem[{{Caproni} {et~al.}(2013){Caproni}, {Abraham}, \& {Monteiro}}]{2013MNRAS.428..280C}
{Caproni}, A., {Abraham}, Z., \& {Monteiro}, H. 2013, \mnras, 428, 280, \dodoi{10.1093/mnras/sts014}

\bibitem[{{Charisi} {et~al.}(2016){Charisi}, {Bartos}, {Haiman}, {Price-Whelan}, {Graham}, {Bellm}, {Laher}, \& {M{\'a}rka}}]{Charisi2016}
{Charisi}, M., {Bartos}, I., {Haiman}, Z., {et~al.} 2016, \mnras, 463, 2145, \dodoi{10.1093/mnras/stw1838}

\bibitem[{{Charisi} {et~al.}(2018){Charisi}, {Haiman}, {Schiminovich}, \& {D'Orazio}}]{2018MNRAS.476.4617C}
{Charisi}, M., {Haiman}, Z., {Schiminovich}, D., \& {D'Orazio}, D.~J. 2018, \mnras, 476, 4617, \dodoi{10.1093/mnras/sty516}

\bibitem[{{Charisi} {et~al.}(2022){Charisi}, {Taylor}, {Runnoe}, {Bogdanovic}, \& {Trump}}]{2022MNRAS.510.5929C}
{Charisi}, M., {Taylor}, S.~R., {Runnoe}, J., {Bogdanovic}, T., \& {Trump}, J.~R. 2022, \mnras, 510, 5929, \dodoi{10.1093/mnras/stab3713}

\bibitem[{{Chen} {et~al.}(2020){Chen}, {Liu}, {Liao}, {Holgado}, {Guo}, {Gruendl}, {Morganson}, {Shen}, {Zhang}, {Abbott}, {Aguena}, {Allam}, {Avila}, {Bertin}, {Bhargava}, {Brooks}, {Burke}, {Carnero Rosell}, {Carollo}, {Carrasco Kind}, {Carretero}, {Costanzi}, {da Costa}, {Davis}, {De Vicente}, {Desai}, {Diehl}, {Doel}, {Everett}, {Flaugher}, {Friedel}, {Frieman}, {Garc{\'\i}a-Bellido}, {Gaztanaga}, {Glazebrook}, {Gruen}, {Gutierrez}, {Hinton}, {Hollowood}, {James}, {Kim}, {Kuehn}, {Kuropatkin}, {Lewis}, {Lidman}, {Lima}, {Maia}, {March}, {Marshall}, {Menanteau}, {Miquel}, {Palmese}, {Paz-Chinch{\'o}n}, {Plazas}, {Sanchez}, {Schubnell}, {Serrano}, {Sevilla-Noarbe}, {Smith}, {Suchyta}, {Swanson}, {Tarle}, {Tucker}, {Norbert Varga}, \& {Walker}}]{Chen2020}
{Chen}, Y.-C., {Liu}, X., {Liao}, W.-T., {et~al.} 2020, \mnras, 499, 2245, \dodoi{10.1093/mnras/staa2957}

\bibitem[{{Chen} {et~al.}(2023){Chen}, {Bao}, {Zhai}, {Fang}, {Hu}, {Du}, {Yang}, {Yao}, {Li}, {Brotherton}, {McLane}, {Zastrocky}, {Olson}, {Bon}, {Bai}, {Fu}, {Liu}, {Wang}, {Maithil}, {Kobulnicky}, {Dale}, {Adelman}, {Caradonna}, {Carter}, {Favro}, {Ferguson}, {Gonzalez}, {Hadding}, {Hagler}, {Murphree}, {Oeur}, {Rogers}, {Roth}, {Schonsberg}, {Stack}, \& {Wang}}]{2023MNRAS.520.1807C}
{Chen}, Y.-J., {Bao}, D.-W., {Zhai}, S., {et~al.} 2023, \mnras, 520, 1807, \dodoi{10.1093/mnras/stad051}

\bibitem[{{Chen} {et~al.}(2024){Chen}, {Zhai}, {Liu}, {Guo}, {Peng}, {Li}, {Songsheng}, {Du}, {Hu}, \& {Wang}}]{Chen_2024_ztf}
{Chen}, Y.-J., {Zhai}, S., {Liu}, J.-R., {et~al.} 2024, \mnras, 527, 12154, \dodoi{10.1093/mnras/stad3981}

\bibitem[{{Cocchiararo} {et~al.}(2024){Cocchiararo}, {Franchini}, {Lupi}, \& {Sesana}}]{2024A&A...691A.250C}
{Cocchiararo}, F., {Franchini}, A., {Lupi}, A., \& {Sesana}, A. 2024, \aap, 691, A250, \dodoi{10.1051/0004-6361/202449598}

\bibitem[{{Dai} {et~al.}(2006){Dai}, {Zhang}, {Xiang}, {Yang}, {Lin}, {Xu}, {Dai}, \& {Zhang}}]{2006IJMPD..15..261D}
{Dai}, B.-Z., {Zhang}, B.-K., {Xiang}, Y., {et~al.} 2006, International Journal of Modern Physics D, 15, 261, \dodoi{10.1142/S0218271806007717}

\bibitem[{{Davis} {et~al.}(2024){Davis}, {Grace}, {Trump}, {Runnoe}, {Henkel}, {Blecha}, {Brandt}, {Casey-Clyde}, {Charisi}, \& {Witt}}]{2024ApJ...965...34D}
{Davis}, M.~C., {Grace}, K.~E., {Trump}, J.~R., {et~al.} 2024, \apj, 965, 34, \dodoi{10.3847/1538-4357/ad276e}

\bibitem[{{De Rosa} {et~al.}(2019){De Rosa}, {Vignali}, {Bogdanovi{\'c}}, {Capelo}, {Charisi}, {Dotti}, {Husemann}, {Lusso}, {Mayer}, {Paragi}, {Runnoe}, {Sesana}, {Steinborn}, {Bianchi}, {Colpi}, {del Valle}, {Frey}, {Gab{\'a}nyi}, {Giustini}, {Guainazzi}, {Haiman}, {Herrera Ruiz}, {Herrero-Illana}, {Iwasawa}, {Komossa}, {Lena}, {Loiseau}, {Perez-Torres}, {Piconcelli}, \& {Volonteri}}]{DeRosa2019}
{De Rosa}, A., {Vignali}, C., {Bogdanovi{\'c}}, T., {et~al.} 2019, \nar, 86, 101525, \dodoi{10.1016/j.newar.2020.101525}

\bibitem[{D'Orazio \& Charisi(2023)}]{Dan&Charisi2023}
D'Orazio, D.~J., \& Charisi, M. 2023, Observational Signatures of Supermassive Black Hole Binaries.
\newblock \doarXiv{2310.16896}

\bibitem[{{D'Orazio} {et~al.}(2024){D'Orazio}, {Duffell}, \& {Tiede}}]{binlite}
{D'Orazio}, D.~J., {Duffell}, P.~C., \& {Tiede}, C. 2024, \apj, 977, 244, \dodoi{10.3847/1538-4357/ad938b}

\bibitem[{{D'Orazio} {et~al.}(2015){D'Orazio}, {Haiman}, \& {Schiminovich}}]{Dorazio2015Nature}
{D'Orazio}, D.~J., {Haiman}, Z., \& {Schiminovich}, D. 2015, \nat, 525, 351, \dodoi{10.1038/nature15262}

\bibitem[{Duffell {et~al.}(2020)Duffell, D'Orazio, Derdzinski, Haiman, MacFadyen, Rosen, \& Zrake}]{duffell_circumbinary_2020}
Duffell, P.~C., D'Orazio, D., Derdzinski, A., {et~al.} 2020, The Astrophysical Journal, 901, 25, \dodoi{10.3847/1538-4357/abab95}

\bibitem[{{EPTA Collaboration} {et~al.}(2023){EPTA Collaboration}, {InPTA Collaboration}, {Antoniadis}, {Arumugam}, {Arumugam}, {Babak}, {Bagchi}, {Bak Nielsen}, {Bassa}, {Bathula}, {Berthereau}, {Bonetti}, {Bortolas}, {Brook}, {Burgay}, {Caballero}, {Chalumeau}, {Champion}, {Chanlaridis}, {Chen}, {Cognard}, {Dandapat}, {Deb}, {Desai}, {Desvignes}, {Dhanda-Batra}, {Dwivedi}, {Falxa}, {Ferdman}, {Franchini}, {Gair}, {Goncharov}, {Gopakumar}, {Graikou}, {Grie{\ss}meier}, {Guillemot}, {Guo}, {Gupta}, {Hisano}, {Hu}, {Iraci}, {Izquierdo-Villalba}, {Jang}, {Jawor}, {Janssen}, {Jessner}, {Joshi}, {Kareem}, {Karuppusamy}, {Keane}, {Keith}, {Kharbanda}, {Kikunaga}, {Kolhe}, {Kramer}, {Krishnakumar}, {Lackeos}, {Lee}, {Liu}, {Liu}, {Lyne}, {McKee}, {Maan}, {Main}, {Mickaliger}, {Ni{\c{t}}u}, {Nobleson}, {Paladi}, {Parthasarathy}, {Perera}, {Perrodin}, {Petiteau}, {Porayko}, {Possenti}, {Prabu}, {Quelquejay Leclere}, {Rana}, {Samajdar}, {Sanidas}, {Sesana}, {Shaifullah}, {Singha}, {Speri}, {Spiewak}, {Srivastava},
  {Stappers}, {Surnis}, {Susarla}, {Susobhanan}, {Takahashi}, {Tarafdar}, {Theureau}, {Tiburzi}, {van der Wateren}, {Vecchio}, {Venkatraman Krishnan}, {Verbiest}, {Wang}, {Wang}, \& {Wu}}]{2023A&A...678A..50E}
{EPTA Collaboration}, {InPTA Collaboration}, {Antoniadis}, J., {et~al.} 2023, \aap, 678, A50, \dodoi{10.1051/0004-6361/202346844}

\bibitem[{{EPTA Collaboration} {et~al.}(2024){EPTA Collaboration}, {InPTA Collaboration}, {Antoniadis}, {Arumugam}, {Arumugam}, {Babak}, {Bagchi}, {Bak Nielsen}, {Bassa}, {Bathula}, {Berthereau}, {Bonetti}, {Bortolas}, {Brook}, {Burgay}, {Caballero}, {Chalumeau}, {Champion}, {Chanlaridis}, {Chen}, {Cognard}, {Dandapat}, {Deb}, {Desai}, {Desvignes}, {Dhanda-Batra}, {Dwivedi}, {Falxa}, {Ferdman}, {Franchini}, {Gair}, {Goncharov}, {Gopakumar}, {Graikou}, {Grie{\ss}meier}, {Gualandris}, {Guillemot}, {Guo}, {Gupta}, {Hisano}, {Hu}, {Iraci}, {Izquierdo-Villalba}, {Jang}, {Jawor}, {Janssen}, {Jessner}, {Joshi}, {Kareem}, {Karuppusamy}, {Keane}, {Keith}, {Kharbanda}, {Kikunaga}, {Kolhe}, {Kramer}, {Krishnakumar}, {Lackeos}, {Lee}, {Liu}, {Liu}, {Lyne}, {McKee}, {Maan}, {Main}, {Mickaliger}, {Ni{\c{t}}u}, {Nobleson}, {Paladi}, {Parthasarathy}, {Perera}, {Perrodin}, {Petiteau}, {Porayko}, {Possenti}, {Prabu}, {Quelquejay Leclere}, {Rana}, {Samajdar}, {Sanidas}, {Sesana}, {Shaifullah}, {Singha}, {Speri}, {Spiewak},
  {Srivastava}, {Stappers}, {Surnis}, {Susarla}, {Susobhanan}, {Takahashi}, {Tarafdar}, {Theureau}, {Tiburzi}, {van der Wateren}, {Vecchio}, {Venkatraman Krishnan}, {Verbiest}, {Wang}, {Wang}, {Wu}, {Auclair}, {Barausse}, {Caprini}, {Crisostomi}, {Fastidio}, {Khizriev}, {Middleton}, {Neronov}, {Postnov}, {Roper Pol}, {Semikoz}, {Smarra}, {Steer}, {Truant}, \& {Valtolina}}]{2024A&A...685A..94E}
---. 2024, \aap, 685, A94, \dodoi{10.1051/0004-6361/202347433}

\bibitem[{{Eracleous} {et~al.}(2012){Eracleous}, {Boroson}, {Halpern}, \& {Liu}}]{eracleous12}
{Eracleous}, M., {Boroson}, T.~A., {Halpern}, J.~P., \& {Liu}, J. 2012, \apjs, 201, 23, \dodoi{10.1088/0067-0049/201/2/23}

\bibitem[{{Graham} {et~al.}(2013){Graham}, {Drake}, {Djorgovski}, {Mahabal}, {Donalek}, {Duan}, \& {Maker}}]{2013MNRAS.434.3423G}
{Graham}, M.~J., {Drake}, A.~J., {Djorgovski}, S.~G., {et~al.} 2013, \mnras, 434, 3423, \dodoi{10.1093/mnras/stt1264}

\bibitem[{{Graham} {et~al.}(2015){Graham}, {Djorgovski}, {Stern}, {Drake}, {Mahabal}, {Donalek}, {Glikman}, {Larson}, \& {Christensen}}]{Graham2015}
{Graham}, M.~J., {Djorgovski}, S.~G., {Stern}, D., {et~al.} 2015, \mnras, 453, 1562, \dodoi{10.1093/mnras/stv1726}

\bibitem[{{Haiman} {et~al.}(2009){Haiman}, {Kocsis}, \& {Menou}}]{2009ApJ...700.1952H}
{Haiman}, Z., {Kocsis}, B., \& {Menou}, K. 2009, \apj, 700, 1952, \dodoi{10.1088/0004-637X/700/2/1952}

\bibitem[{{Hu} {et~al.}(2020){Hu}, {D'Orazio}, {Haiman}, {Smith}, {Snios}, {Charisi}, \& {Di Stefano}}]{2020MNRAS.495.4061H}
{Hu}, B.~X., {D'Orazio}, D.~J., {Haiman}, Z., {et~al.} 2020, \mnras, 495, 4061, \dodoi{10.1093/mnras/staa1312}

\bibitem[{{Huijse} {et~al.}(2025){Huijse}, {Davelaar}, {De Ridder}, {Jannsen}, \& {Aerts}}]{2025arXiv250516884H}
{Huijse}, P., {Davelaar}, J., {De Ridder}, J., {Jannsen}, N., \& {Aerts}, C. 2025, arXiv e-prints, arXiv:2505.16884, \dodoi{10.48550/arXiv.2505.16884}

\bibitem[{{Ivezi{\'c}} {et~al.}(2014){Ivezi{\'c}}, {Connolly}, {Vanderplas}, \& {Gray}}]{astroMLText}
{Ivezi{\'c}}, {\v Z}., {Connolly}, A., {Vanderplas}, J., \& {Gray}, A. 2014, Statistics, Data Mining and Machine Learning in Astronomy (Princeton University Press)

\bibitem[{{Ivezi{\'c}} {et~al.}(2019){Ivezi{\'c}}, {Kahn}, {Tyson}, {Abel}, {Acosta}, {Allsman}, {Alonso}, {AlSayyad}, {Anderson}, {Andrew}, {Angel}, {Angeli}, {Ansari}, {Antilogus}, {Araujo}, {Armstrong}, {Arndt}, {Astier}, {Aubourg}, {Auza}, {Axelrod}, {Bard}, {Barr}, {Barrau}, {Bartlett}, {Bauer}, {Bauman}, {Baumont}, {Bechtol}, {Bechtol}, {Becker}, {Becla}, {Beldica}, {Bellavia}, {Bianco}, {Biswas}, {Blanc}, {Blazek}, {Blandford}, {Bloom}, {Bogart}, {Bond}, {Booth}, {Borgland}, {Borne}, {Bosch}, {Boutigny}, {Brackett}, {Bradshaw}, {Brandt}, {Brown}, {Bullock}, {Burchat}, {Burke}, {Cagnoli}, {Calabrese}, {Callahan}, {Callen}, {Carlin}, {Carlson}, {Chandrasekharan}, {Charles-Emerson}, {Chesley}, {Cheu}, {Chiang}, {Chiang}, {Chirino}, {Chow}, {Ciardi}, {Claver}, {Cohen-Tanugi}, {Cockrum}, {Coles}, {Connolly}, {Cook}, {Cooray}, {Covey}, {Cribbs}, {Cui}, {Cutri}, {Daly}, {Daniel}, {Daruich}, {Daubard}, {Daues}, {Dawson}, {Delgado}, {Dellapenna}, {de Peyster}, {de Val-Borro}, {Digel}, {Doherty}, {Dubois},
  {Dubois-Felsmann}, {Durech}, {Economou}, {Eifler}, {Eracleous}, {Emmons}, {Fausti Neto}, {Ferguson}, {Figueroa}, {Fisher-Levine}, {Focke}, {Foss}, {Frank}, {Freemon}, {Gangler}, {Gawiser}, {Geary}, {Gee}, {Geha}, {Gessner}, {Gibson}, {Gilmore}, {Glanzman}, {Glick}, {Goldina}, {Goldstein}, {Goodenow}, {Graham}, {Gressler}, {Gris}, {Guy}, {Guyonnet}, {Haller}, {Harris}, {Hascall}, {Haupt}, {Hernandez}, {Herrmann}, {Hileman}, {Hoblitt}, {Hodgson}, {Hogan}, {Howard}, {Huang}, {Huffer}, {Ingraham}, {Innes}, {Jacoby}, {Jain}, {Jammes}, {Jee}, {Jenness}, {Jernigan}, {Jevremovi{\'c}}, {Johns}, {Johnson}, {Johnson}, {Jones}, {Juramy-Gilles}, {Juri{\'c}}, {Kalirai}, {Kallivayalil}, {Kalmbach}, {Kantor}, {Karst}, {Kasliwal}, {Kelly}, {Kessler}, {Kinnison}, {Kirkby}, {Knox}, {Kotov}, {Krabbendam}, {Krughoff}, {Kub{\'a}nek}, {Kuczewski}, {Kulkarni}, {Ku}, {Kurita}, {Lage}, {Lambert}, {Lange}, {Langton}, {Le Guillou}, {Levine}, {Liang}, {Lim}, {Lintott}, {Long}, {Lopez}, {Lotz}, {Lupton}, {Lust}, {MacArthur}, {Mahabal},
  {Mandelbaum}, {Markiewicz}, {Marsh}, {Marshall}, {Marshall}, {May}, {McKercher}, {McQueen}, {Meyers}, {Migliore}, {Miller}, {Mills}, {Miraval}, {Moeyens}, {Moolekamp}, {Monet}, {Moniez}, {Monkewitz}, {Montgomery}, {Morrison}, {Mueller}, {Muller}, {Mu{\~n}oz Arancibia}, {Neill}, {Newbry}, {Nief}, {Nomerotski}, {Nordby}, {O'Connor}, {Oliver}, {Olivier}, {Olsen}, {O'Mullane}, {Ortiz}, {Osier}, {Owen}, {Pain}, {Palecek}, {Parejko}, {Parsons}, {Pease}, {Peterson}, {Peterson}, {Petravick}, {Libby Petrick}, {Petry}, {Pierfederici}, {Pietrowicz}, {Pike}, {Pinto}, {Plante}, {Plate}, {Plutchak}, {Price}, {Prouza}, {Radeka}, {Rajagopal}, {Rasmussen}, {Regnault}, {Reil}, {Reiss}, {Reuter}, {Ridgway}, {Riot}, {Ritz}, {Robinson}, {Roby}, {Roodman}, {Rosing}, {Roucelle}, {Rumore}, {Russo}, {Saha}, {Sassolas}, {Schalk}, {Schellart}, {Schindler}, {Schmidt}, {Schneider}, {Schneider}, {Schoening}, {Schumacher}, {Schwamb}, {Sebag}, {Selvy}, {Sembroski}, {Seppala}, {Serio}, {Serrano}, {Shaw}, {Shipsey}, {Sick}, {Silvestri},
  {Slater}, {Smith}, {Smith}, {Sobhani}, {Soldahl}, {Storrie-Lombardi}, {Stover}, {Strauss}, {Street}, {Stubbs}, {Sullivan}, {Sweeney}, {Swinbank}, {Szalay}, {Takacs}, {Tether}, {Thaler}, {Thayer}, {Thomas}, {Thornton}, {Thukral}, {Tice}, {Trilling}, {Turri}, {Van Berg}, {Vanden Berk}, {Vetter}, {Virieux}, {Vucina}, {Wahl}, {Walkowicz}, {Walsh}, {Walter}, {Wang}, {Wang}, {Warner}, {Wiecha}, {Willman}, {Winters}, {Wittman}, {Wolff}, {Wood-Vasey}, {Wu}, {Xin}, {Yoachim}, \& {Zhan}}]{2019ApJ...873..111I}
{Ivezi{\'c}}, {\v{Z}}., {Kahn}, S.~M., {Tyson}, J.~A., {et~al.} 2019, \apj, 873, 111, \dodoi{10.3847/1538-4357/ab042c}

\bibitem[{{Kasliwal} {et~al.}(2015){Kasliwal}, {Vogeley}, \& {Richards}}]{2015MNRAS.451.4328K}
{Kasliwal}, V.~P., {Vogeley}, M.~S., \& {Richards}, G.~T. 2015, \mnras, 451, 4328, \dodoi{10.1093/mnras/stv1230}

\bibitem[{{Kaspi} {et~al.}(2021){Kaspi}, {Brandt}, {Maoz}, {Netzer}, {Schneider}, {Shemmer}, \& {Grier}}]{2021arXiv210600691K}
{Kaspi}, S., {Brandt}, W.~N., {Maoz}, D., {et~al.} 2021, arXiv e-prints, arXiv:2106.00691, \dodoi{10.48550/arXiv.2106.00691}

\bibitem[{{Kauffmann} \& {Haehnelt}(2000)}]{KauffmanHaehnelt2000}
{Kauffmann}, G., \& {Haehnelt}, M. 2000, \mnras, 311, 576, \dodoi{10.1046/j.1365-8711.2000.03077.x}

\bibitem[{{Kelley} {et~al.}(2019){Kelley}, {Charisi}, {Burke-Spolaor}, {Simon}, {Blecha}, {Bogdanovic}, {Colpi}, {Comerford}, {D'Orazio}, {Dotti}, {Eracleous}, {Graham}, {Greene}, {Haiman}, {Holley-Bockelmann}, {Kara}, {Kelly}, {Komossa}, {Larson}, {Liu}, {Ma}, {Noble}, {Paschalidis}, {Rafikov}, {Ravi}, {Runnoe}, {Sesana}, {Stern}, {Strauss}, {U}, {Volonteri}, \& {Nanograv Collaboration}}]{2019BAAS...51c.490K}
{Kelley}, L., {Charisi}, M., {Burke-Spolaor}, S., {et~al.} 2019, \baas, 51, 490.
\newblock \doarXiv{1903.07644}

\bibitem[{{Kelley} {et~al.}(2018){Kelley}, {Blecha}, {Hernquist}, {Sesana}, \& {Taylor}}]{Kelley2018}
{Kelley}, L.~Z., {Blecha}, L., {Hernquist}, L., {Sesana}, A., \& {Taylor}, S.~R. 2018, \mnras, 477, 964, \dodoi{10.1093/mnras/sty689}

\bibitem[{{Kelley} {et~al.}(2021){Kelley}, {D'Orazio}, \& {Di Stefano}}]{2021MNRAS.508.2524K}
{Kelley}, L.~Z., {D'Orazio}, D.~J., \& {Di Stefano}, R. 2021, \mnras, 508, 2524, \dodoi{10.1093/mnras/stab2776}

\bibitem[{{Kelly} {et~al.}(2009){Kelly}, {Bechtold}, \& {Siemiginowska}}]{Kelly2009}
{Kelly}, B.~C., {Bechtold}, J., \& {Siemiginowska}, A. 2009, \apj, 698, 895, \dodoi{10.1088/0004-637X/698/1/895}

\bibitem[{{Kelly} {et~al.}(2014){Kelly}, {Becker}, {Sobolewska}, {Siemiginowska}, \& {Uttley}}]{2014ApJ...788...33K}
{Kelly}, B.~C., {Becker}, A.~C., {Sobolewska}, M., {Siemiginowska}, A., \& {Uttley}, P. 2014, \apj, 788, 33, \dodoi{10.1088/0004-637X/788/1/33}

\bibitem[{{Kocsis} {et~al.}(2006){Kocsis}, {Frei}, {Haiman}, \& {Menou}}]{Kocsis+2006}
{Kocsis}, B., {Frei}, Z., {Haiman}, Z., \& {Menou}, K. 2006, \apj, 637, 27, \dodoi{10.1086/498236}

\bibitem[{{Koen}(2021)}]{Koen_2021}
{Koen}, C. 2021, \aj, 161, 281, \dodoi{10.3847/1538-3881/abf64e}

\bibitem[{{Kormendy} \& {Ho}(2013)}]{KormendyHo2013}
{Kormendy}, J., \& {Ho}, L.~C. 2013, \araa, 51, 511, \dodoi{10.1146/annurev-astro-082708-101811}

\bibitem[{{Koz{\l}owski} {et~al.}(2010){Koz{\l}owski}, {Kochanek}, {Udalski}, {Wyrzykowski}, {Soszy{\'n}ski}, {Szyma{\'n}ski}, {Kubiak}, {Pietrzy{\'n}ski}, {Szewczyk}, {Ulaczyk}, {Poleski}, \& {OGLE Collaboration}}]{Koz2010}
{Koz{\l}owski}, S., {Kochanek}, C.~S., {Udalski}, A., {et~al.} 2010, \apj, 708, 927, \dodoi{10.1088/0004-637X/708/2/927}

\bibitem[{{Law} {et~al.}(2009){Law}, {Kulkarni}, {Dekany}, {Ofek}, {Quimby}, {Nugent}, {Surace}, {Grillmair}, {Bloom}, {Kasliwal}, {Bildsten}, {Brown}, {Cenko}, {Ciardi}, {Croner}, {Djorgovski}, {van Eyken}, {Filippenko}, {Fox}, {Gal-Yam}, {Hale}, {Hamam}, {Helou}, {Henning}, {Howell}, {Jacobsen}, {Laher}, {Mattingly}, {McKenna}, {Pickles}, {Poznanski}, {Rahmer}, {Rau}, {Rosing}, {Shara}, {Smith}, {Starr}, {Sullivan}, {Velur}, {Walters}, \& {Zolkower}}]{Law2009}
{Law}, N.~M., {Kulkarni}, S.~R., {Dekany}, R.~G., {et~al.} 2009, \pasp, 121, 1395, \dodoi{10.1086/648598}

\bibitem[{{Li} {et~al.}(2019){Li}, {Wang}, {Zhang}, {Wang}, {Huang}, {Lu}, {Hu}, {Du}, {Bon}, {Ho}, {Bai}, {Bian}, {Yuan}, {Winkler}, {Denissyuk}, {Valiullin}, {Bon}, \& {Popovi{\'c}}}]{2019ApJS..241...33L}
{Li}, Y.-R., {Wang}, J.-M., {Zhang}, Z.-X., {et~al.} 2019, \apjs, 241, 33, \dodoi{10.3847/1538-4365/ab0ec5}

\bibitem[{{Liu} {et~al.}(2019){Liu}, {Gezari}, {Ayers}, {Burgett}, {Chambers}, {Hodapp}, {Huber}, {Kudritzki}, {Metcalfe}, {Tonry}, {Wainscoat}, \& {Waters}}]{2019ApJ...884...36L}
{Liu}, T., {Gezari}, S., {Ayers}, M., {et~al.} 2019, \apj, 884, 36, \dodoi{10.3847/1538-4357/ab40cb}

\bibitem[{Lomb(1976)}]{lomb_least-squares_1976}
Lomb, N.~R. 1976, Astrophysics and Space Science, 39, 447, \dodoi{10.1007/BF00648343}

\bibitem[{{Luo} {et~al.}(2025){Luo}, {Jiang}, \& {Liu}}]{2025ApJ...978...86L}
{Luo}, D., {Jiang}, N., \& {Liu}, X. 2025, \apj, 978, 86, \dodoi{10.3847/1538-4357/ad9245}

\bibitem[{{MacLeod} {et~al.}(2010){MacLeod}, {Ivezi{\'c}}, {Kochanek}, {Koz{\l}owski}, {Kelly}, {Bullock}, {Kimball}, {Sesar}, {Westman}, {Brooks}, {Gibson}, {Becker}, \& {de Vries}}]{MacLeod+2010}
{MacLeod}, C.~L., {Ivezi{\'c}}, {\v Z}., {Kochanek}, C.~S., {et~al.} 2010, \apj, 721, 1014, \dodoi{10.1088/0004-637X/721/2/1014}

\bibitem[{{Miles} {et~al.}(2025){Miles}, {Shannon}, {Reardon}, {Bailes}, {Champion}, {Geyer}, {Gitika}, {Grunthal}, {Keith}, {Kramer}, {Kulkarni}, {Nathan}, {Parthasarathy}, {Singha}, {Theureau}, {Thrane}, {Abbate}, {Buchner}, {Cameron}, {Camilo}, {Moreschi}, {Shaifullah}, {Shamohammadi}, {Possenti}, \& {Krishnan}}]{MPTA}
{Miles}, M.~T., {Shannon}, R.~M., {Reardon}, D.~J., {et~al.} 2025, \mnras, 536, 1489, \dodoi{10.1093/mnras/stae2571}

\bibitem[{{Moreno} {et~al.}(2019){Moreno}, {Vogeley}, {Richards}, \& {Yu}}]{2019PASP..131f3001M}
{Moreno}, J., {Vogeley}, M.~S., {Richards}, G.~T., \& {Yu}, W. 2019, \pasp, 131, 063001, \dodoi{10.1088/1538-3873/ab1597}

\bibitem[{{Peterson} {et~al.}(2002){Peterson}, {Berlind}, {Bertram}, {Bischoff}, {Bochkarev}, {Borisov}, {Burenkov}, {Calkins}, {Carrasco}, {Chavushyan}, {Chornock}, {Dietrich}, {Doroshenko}, {Ezhkova}, {Filippenko}, {Gilbert}, {Huchra}, {Kollatschny}, {Leonard}, {Li}, {Lyuty}, {Malkov}, {Matheson}, {Merkulova}, {Mikhailov}, {Modjaz}, {Onken}, {Pogge}, {Pronik}, {Qian}, {Romano}, {Sergeev}, {Sergeeva}, {Shapovalova}, {Spiridonova}, {Tao}, {Tokarz}, {Valdes}, {Vlasiuk}, {Wagner}, \& {Wilkes}}]{2002ApJ...581..197P}
{Peterson}, B.~M., {Berlind}, P., {Bertram}, R., {et~al.} 2002, \apj, 581, 197, \dodoi{10.1086/344197}

\bibitem[{{Rau} {et~al.}(2009){Rau}, {Kulkarni}, {Law}, {Bloom}, {Ciardi}, {Djorgovski}, {Fox}, {Gal-Yam}, {Grillmair}, {Kasliwal}, {Nugent}, {Ofek}, {Quimby}, {Reach}, {Shara}, {Bildsten}, {Cenko}, {Drake}, {Filippenko}, {Helfand}, {Helou}, {Howell}, {Poznanski}, \& {Sullivan}}]{Rau2009}
{Rau}, A., {Kulkarni}, S.~R., {Law}, N.~M., {et~al.} 2009, \pasp, 121, 1334, \dodoi{10.1086/605911}

\bibitem[{{Reardon} {et~al.}(2023){Reardon}, {Zic}, {Shannon}, {Hobbs}, {Bailes}, {Di Marco}, {Kapur}, {Rogers}, {Thrane}, {Askew}, {Bhat}, {Cameron}, {Cury{\l}o}, {Coles}, {Dai}, {Goncharov}, {Kerr}, {Kulkarni}, {Levin}, {Lower}, {Manchester}, {Mandow}, {Miles}, {Nathan}, {Os{\l}owski}, {Russell}, {Spiewak}, {Zhang}, \& {Zhu}}]{2023ApJ...951L...6R}
{Reardon}, D.~J., {Zic}, A., {Shannon}, R.~M., {et~al.} 2023, \apjl, 951, L6, \dodoi{10.3847/2041-8213/acdd02}

\bibitem[{{Richstone} {et~al.}(1998){Richstone}, {Ajhar}, {Bender}, {Bower}, {Dressler}, {Faber}, {Filippenko}, {Gebhardt}, {Green}, {Ho}, {Kormendy}, {Lauer}, {Magorrian}, \& {Tremaine}}]{Richstone1998}
{Richstone}, D., {Ajhar}, E.~A., {Bender}, R., {et~al.} 1998, \nat, 385, A14, \dodoi{10.48550/arXiv.astro-ph/9810378}

\bibitem[{Robnik {et~al.}(2024)Robnik, Bayer, Charisi, Haiman, Lin, \& Seljak}]{Robnik2024}
Robnik, J., Bayer, A.~E., Charisi, M., {et~al.} 2024, Monthly Notices of the Royal Astronomical Society, 534, 1609, \dodoi{10.1093/mnras/stae2220}

\bibitem[{{Roos} {et~al.}(1993){Roos}, {Kaastra}, \& {Hummel}}]{1993ApJ...409..130R}
{Roos}, N., {Kaastra}, J.~S., \& {Hummel}, C.~A. 1993, \apj, 409, 130, \dodoi{10.1086/172647}

\bibitem[{{Saade} {et~al.}(2020){Saade}, {Stern}, {Brightman}, {Haiman}, {Djorgovski}, {D'Orazio}, {Ford}, {Graham}, {Jun}, {Kraft}, {McKernan}, {Vikhlinin}, \& {Walton}}]{2020ApJ...900..148S}
{Saade}, M.~L., {Stern}, D., {Brightman}, M., {et~al.} 2020, \apj, 900, 148, \dodoi{10.3847/1538-4357/abad31}

\bibitem[{Scargle(1982)}]{scargle_studies_1982}
Scargle, J.~D. 1982, The Astrophysical Journal, 263, 835, \dodoi{10.1086/160554}

\bibitem[{{Smith} {et~al.}(2018){Smith}, {Mushotzky}, {Boyd}, {Malkan}, {Howell}, \& {Gelino}}]{2018ApJ...857..141S}
{Smith}, K.~L., {Mushotzky}, R.~F., {Boyd}, P.~T., {et~al.} 2018, \apj, 857, 141, \dodoi{10.3847/1538-4357/aab88d}

\bibitem[{{Suberlak} {et~al.}(2021){Suberlak}, {Ivezi{\'c}}, \& {MacLeod}}]{2021ApJ...907...96S}
{Suberlak}, K.~L., {Ivezi{\'c}}, {\v{Z}}., \& {MacLeod}, C. 2021, \apj, 907, 96, \dodoi{10.3847/1538-4357/abc698}

\bibitem[{{S{\"u}veges} {et~al.}(2015){S{\"u}veges}, {Guy}, {Eyer}, {Cuypers}, {Holl}, {Lecoeur-Ta{\"\i}bi}, {Mowlavi}, {Nienartowicz}, {Blanco}, {Rimoldini}, \& {Ruiz}}]{Suveges_2015}
{S{\"u}veges}, M., {Guy}, L.~P., {Eyer}, L., {et~al.} 2015, \mnras, 450, 2052, \dodoi{10.1093/mnras/stv719}

\bibitem[{{Taylor}(2021)}]{2021arXiv210513270T}
{Taylor}, S.~R. 2021, arXiv e-prints, arXiv:2105.13270.
\newblock \doarXiv{2105.13270}

\bibitem[{{Taylor} {et~al.}(2020){Taylor}, {van Haasteren}, \& {Sesana}}]{2020PhRvD.102h4039T}
{Taylor}, S.~R., {van Haasteren}, R., \& {Sesana}, A. 2020, \prd, 102, 084039, \dodoi{10.1103/PhysRevD.102.084039}

\bibitem[{{Timmer} \& {K{\"o}nig}(1995)}]{Timmer1995}
{Timmer}, J., \& {K{\"o}nig}, M. 1995, \aap, 300, 707

\bibitem[{VanderPlas(2018)}]{vanderplas_understanding_2018}
VanderPlas, J.~T. 2018, The Astrophysical Journal Supplement Series, 236, 16, \dodoi{10.3847/1538-4365/aab766}

\bibitem[{{Vaughan} {et~al.}(2016){Vaughan}, {Uttley}, {Markowitz}, {Huppenkothen}, {Middleton}, {Alston}, {Scargle}, \& {Farr}}]{Vaughan2016}
{Vaughan}, S., {Uttley}, P., {Markowitz}, A.~G., {et~al.} 2016, \mnras, 461, 3145, \dodoi{10.1093/mnras/stw1412}

\bibitem[{Westernacher-Schneider {et~al.}(2022)Westernacher-Schneider, Zrake, MacFadyen, \& Haiman}]{westernacher-schneider_multiband_2022}
Westernacher-Schneider, J.~R., Zrake, J., MacFadyen, A., \& Haiman, Z. 2022, Physical Review D, 106, 103010, \dodoi{10.1103/PhysRevD.106.103010}

\bibitem[{{Witt} {et~al.}(2022){Witt}, {Charisi}, {Taylor}, \& {Burke-Spolaor}}]{Witt2022}
{Witt}, C.~A., {Charisi}, M., {Taylor}, S.~R., \& {Burke-Spolaor}, S. 2022, \apj, 936, 89, \dodoi{10.3847/1538-4357/ac8356}

\bibitem[{{Xin} {et~al.}(2020){Xin}, {Charisi}, {Haiman}, {Schiminovich}, {Graham}, {Stern}, \& {D'Orazio}}]{2020MNRAS.496.1683X}
{Xin}, C., {Charisi}, M., {Haiman}, Z., {et~al.} 2020, \mnras, 496, 1683, \dodoi{10.1093/mnras/staa1643}

\bibitem[{{Xin} \& {Haiman}(2021)}]{2021MNRAS.506.2408X}
{Xin}, C., \& {Haiman}, Z. 2021, \mnras, 506, 2408, \dodoi{10.1093/mnras/stab1856}

\bibitem[{{Xu} {et~al.}(2023){Xu}, {Chen}, {Guo}, {Jiang}, {Wang}, {Xu}, {Xue}, {Nicolas Caballero}, {Yuan}, {Xu}, {Wang}, {Hao}, {Luo}, {Lee}, {Han}, {Jiang}, {Shen}, {Wang}, {Wang}, {Xu}, {Wu}, {Manchester}, {Qian}, {Guan}, {Huang}, {Sun}, \& {Zhu}}]{2023RAA....23g5024X}
{Xu}, H., {Chen}, S., {Guo}, Y., {et~al.} 2023, Research in Astronomy and Astrophysics, 23, 075024, \dodoi{10.1088/1674-4527/acdfa5}

\bibitem[{{Yu} {et~al.}(2022){Yu}, {Richards}, {Vogeley}, {Moreno}, \& {Graham}}]{2022ApJ...936..132Y}
{Yu}, W., {Richards}, G.~T., {Vogeley}, M.~S., {Moreno}, J., \& {Graham}, M.~J. 2022, \apj, 936, 132, \dodoi{10.3847/1538-4357/ac8351}

\bibitem[{{Zrake} {et~al.}(2021){Zrake}, {Tiede}, {MacFadyen}, \& {Haiman}}]{2021ApJ...909L..13Z}
{Zrake}, J., {Tiede}, C., {MacFadyen}, A., \& {Haiman}, Z. 2021, \apjl, 909, L13, \dodoi{10.3847/2041-8213/abdd1c}

\bibitem[{{Zu} {et~al.}(2013){Zu}, {Kochanek}, {Koz{\l}owski}, \& {Udalski}}]{Javelin}
{Zu}, Y., {Kochanek}, C.~S., {Koz{\l}owski}, S., \& {Udalski}, A. 2013, \apj, 765, 106, \dodoi{10.1088/0004-637X/765/2/106}

\end{thebibliography}
\label{lastpage}
\end{document}